%% file: main.tex
\documentclass[journal,twocolumn, 10pt]{IEEEtran}
 \pdfoutput=1
\usepackage{epsfig,makeidx,color,epstopdf}
\usepackage{amsmath,amssymb,amsthm,bbm}
\usepackage{cite,graphicx}
\usepackage{enumerate}
\usepackage{epstopdf}
\usepackage{cite}
\usepackage{amsfonts}
\usepackage{mathtools, cuted}
\usepackage{lipsum}
\usepackage{indentfirst}
\usepackage{subfig}
\usepackage[font=footnotesize]{caption} 
\usepackage{csquotes}
\usepackage{float}
\usepackage{comment}
\DeclareMathOperator\erf{erf}
\makeatletter
\def\footnoterule{\kern-3\p@
  \hrule \@width 2in \kern 2.6\p@} 
\makeatother
   \usepackage[table]{xcolor}
   \usepackage{stfloats}
   \usepackage{cases}

\usepackage{array}
\usepackage{dsfont}
\newcolumntype{P}[1]{>{\centering\arraybackslash}p{#1}}
\newcolumntype{M}[1]{>{\centering\arraybackslash}m{#1}}

\DeclareMathAlphabet{\mathpzc}{OT1}{pzc}{m}{it}
\def\defh{
\mbox{\footnotesize $ \begin{array}{c} H_0 \cr > \cr < \cr H_1 \end{array} $}}

\def\be{ \begin{equation} }
\def\ee{ \end{equation} }
\def\bea{ \begin{eqnarray} }
\def\eea{ \end{eqnarray} }

\def\b0{{\bf 0}}

\def\cS{{\cal S}}

\ifCLASSOPTIONonecolumn
\else
  
\fi

\makeatletter
\newcommand*\dashline{\rotatebox[origin=c]{90}{$\dabar@\dabar@\dabar@$}}
\makeatother
\makeatletter
\newcommand*{\rom}[1]{\expandafter\@slowromancap\romannumeral #1@}
\makeatother

\begin{document}
\title{Subcarrier-wise Backscatter Communications \\ over Ambient OFDM for Low Power IoT}
\author{Mahyar Nemati, Morteza Soltani, Jie Ding, and Jinho Choi
\thanks{M. Nemati, J. Ding, and J. Choi are with
the School of Information Technology, 
Deakin University, 
Geelong, VIC 3220, Australia (e-mail: nematim@deakin.edu.au, yxdj2010@gmail.com, jinho.choi@deakin.edu.au)}
\thanks{M. Soltani is with the Department of Electrical and Computer Engineering, University of Idaho, 83844 Moscow, Idaho, USA (e-mail: msoltani@uidaho.edu) }}
\date{today}
\maketitle

\begin{abstract}
Ambient backscatter communication (AmBC) over orthogonal-frequency-division-multiplexing (OFDM) signals has recently been proposed as an appealing technique for low power Internet-of-Things (IoT) applications. The special spectrum structure of OFDM signals provides a range of flexibility in terms of bit-error-rate (BER) performance, data rate, and power consumption. In this paper, we study subcarrier-wise backscatter communication over ambient OFDM signals.
This new AmBC is to exploit the special spectrum structure of OFDM to transmit data over its squeezed orthogonal subcarriers. We propose a basis transmission scheme and its two modifications to support a higher data rate with superior BER performance compared to existing methods. The basis scheme can transmit one bit per subcarrier using on-off keying (OOK) modulation in the frequency domain.
 In the first modification, interleaved subcarrier block transmission model is employed to improve the BER performance of the system in frequency-selective channels. It results in a trade-off between the size of the blocks and data rate. 
 Thus, in the second modification, interleaved index modulation (IM) is employed to mitigate the data rate decrementation of the former modification. It also stabilizes and controls the power of the signal to result in interference reduction for a legacy receiver.
 Analytical and numerical evaluations provide a proof to see the performance of the proposed method in terms of BER, data rate, and interference.
\end{abstract}
{\IEEEkeywords
AmBC, data rate, interference, OFDM, power, subcarrier.}

\vspace{-6mm}
\section{Introduction}
\input{Sections/1_introduction.tex}

\section{System Model}
\input{Sections/2_system.tex}

\section{Signal Detection and BER analysis}
\input{Sections/3_detection.tex}

\section{Description of Modifications}
\input{Sections/4_modifs.tex}

\section{Numerical Results}
\input{Sections/5_simul.tex}

\section{Conclusions}
\input{Sections/Conclusions.tex}

\section{Acknowledgement}
This work was supported by Australian Research Council (ARC) Discovery 2020 Funding, under grant number DP200100391.

\begin{appendices}

\section{}
\input{Subs/Ap1.tex}
\label{AP1}

\section{}
\input{Subs/AP2.tex}
\label{AP2}

\end{appendices}

\bibliographystyle{IEEEtran}
\bibliography{ref}

\end{document}

%% file: Sections/1_introduction.tex
Ambient backscatter communication (AmBC) \cite{liu2013} is a promising technology for low power Internet-of-Things (IoT) applications, e.g., green IoT\cite{GRN}, which is also considered to be a type of joint power-information transmission technologies. 
AmBC technology enables small battery-less IoT devices (e.g., passive tags) to harvest power from far-field radio-frequency (RF) signals. Then, a new communication takes place over the same spectrum of the RF signals \cite{noma}. It makes AmBC a promising power- and spectrum-efficient technology for future IoT. In a nutshell, battery-less IoT devices or tags modulate and backscatter the surrounding RF signals towards a receiver (i.e., reader)\cite{van2018,liu2013}. Different from the conventional backscatter communications, such as in RF identification (RFID) systems \cite{RFIDboyer,rev,rfid,semi,Semi1}, AmBC does not require a dedicated source or reader for direct device-to-device (D2D) and even multi-hop communications \cite{RFIDint,of-3}. 

\begin{figure}[t]
\centering
\captionsetup{width=1\linewidth}
  \includegraphics[width=7.5cm, height=4.5cm]{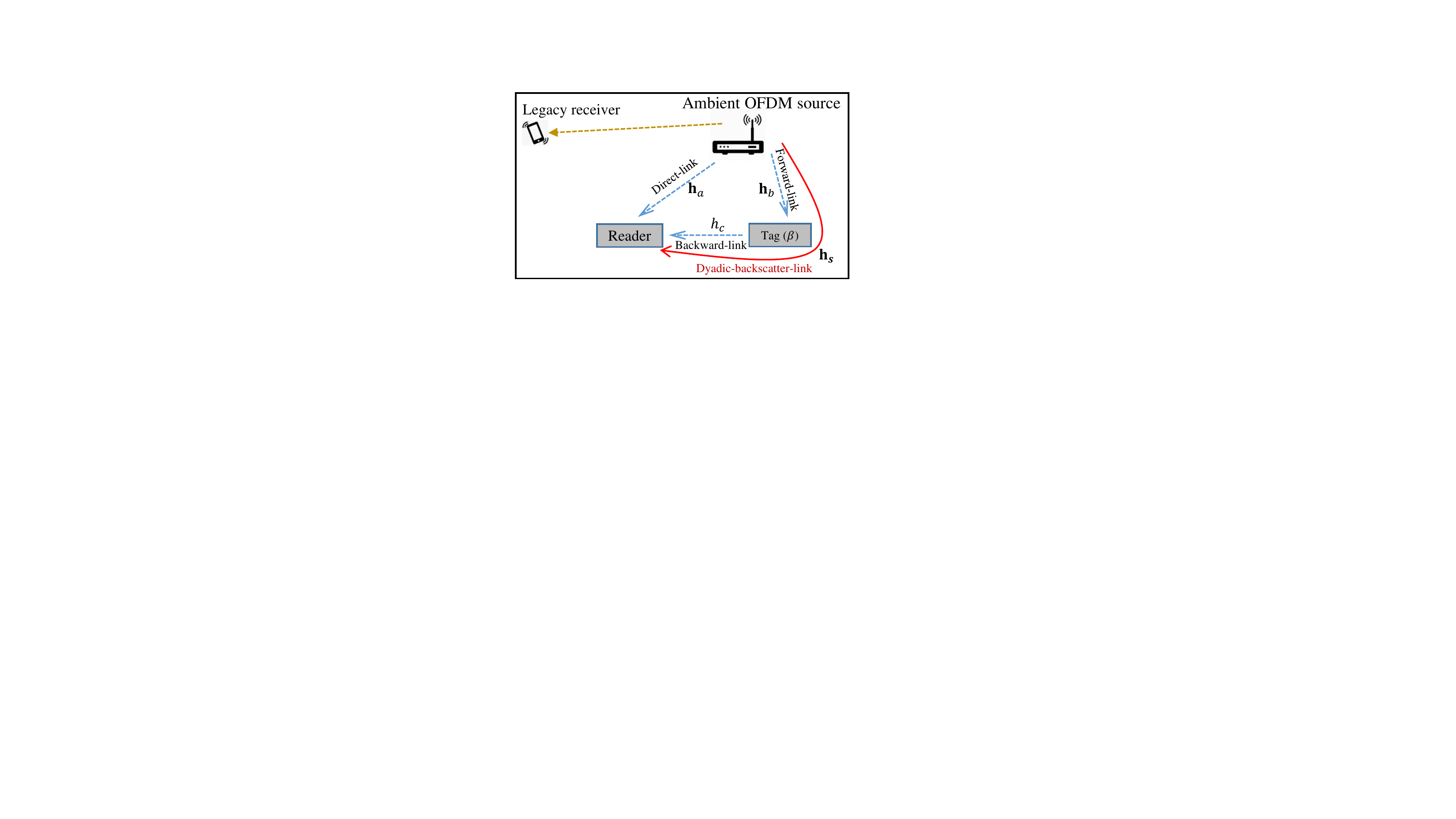}%
  \caption{Illustration of AmBC system model.}%
  \label{fig:sys}
\end{figure}

In general, an AmBC system includes a tag, a reader, an ambient RF source, and a legacy receiver as shown in Fig. \ref{fig:sys}. A source broadcasts its RF signal to communicate with the legacy receiver. The tag also receives the RF signal through the forward-link channel due to its relatively short distance from the source. Integrated circuit of the tag includes passive components to harvest/capture enough power from ambient RF signal to modulate and backscatter it towards the reader. The reader also receives the ambient RF signal which is considered to be direct-link interference for signal detection. In addition, AmBC can cause an interference for a legacy receiver due to same spectrum utilization \cite{noma,of-3}.
It is also noteworthy that there are two other types of tags such as active and semi-passive tags \cite{rfid,semi,Semi1} in conventional backscattering systems (e.g., RFID systems, wireless sensor networks (WSN)); however, in this study, we only focus on passive tags to comply with what has been mostly considered in the literature thus far, and also to emphasize the suitability of the AmBC for low power IoT applications. Nevertheless, it does not restrict our AmBC system model to only passive tags.

\label{sec1}
In the literature, a number of approaches are studied for AmBC over different dominant RF signals transmitted from various sources such as TV/FM towers, cellular base stations, and even Wi-Fi access points (APs). AmBC was firstly set up over ambient digital TV (DTV) signals for a short-range (up to $8$ m) D2D communication with a data rate of $1$ kbps \cite{liu2013}. Likewise, the AmBC system in \cite{chan2} with the same principles achieved a higher data rate of $20$ kbps using Wi-Fi signals. Subsequently, a data rate of $1$ Mbps was obtained in \cite{backfi} by using the principles of full-duplex systems \cite{of-3}. Afterwards, a noncoherent detection method using multiple antennas was proposed in \cite{wang2016} to remove the need of channel state information (CSI) for signal detection in AmBC systems regardless of the signal type. Furthermore, some experiments were carried out in \cite{farm1} for AmBC over FM signals. It supported a low data transmission rate of up to $500$ bps for agricultural monitoring sensors. However, the front-end of these studies only allows data transmission in one single band which makes them difficult to work in different environments \cite{BPfilter}. Also, none of the aforementioned approaches evaluated or exploited the physical layer structure of the signals/waveforms for performance improvements. 

Among different dominant RF signals/waveforms, orthogonal-frequency-division-multiplexing (OFDM) waveform has a special spectrum structure which can provide a range of flexibility in terms of bit-error-rate (BER) performance, data rate, and power consumption. Moreover, it is a common modulation scheme in most of the modern communication systems (e.g., LTE/LTE-A, Wi-Fi, and even first commercial 5th generation (5G) of mobile communication standard (3GPP Release 15)\cite{5G3gpp}). 
In \cite{of-3,Jin_OF,of-0,of,Sensor,rev3}, three different AmBC methods were proposed to transmit only one bit over one OFDM symbol (duration). Authors in \cite{of-3} showed that backscattering one bit over one OFDM symbol provides a satisfactory data rate ($\sim 1$ Mbps using OFDM system based on IEEE 802.11a) for short distances ($1-3$ m). 
A method in \cite{Jin_OF} used matched-filtering at the tag to impose a certain property that improves the signal detection performance at the reader. In \cite{of-0,of,Sensor}, an AmBC approach was evaluated which exploits the cyclic prefix (CP) of OFDM signals to address direct-link interference cancellation when the duration of the CP, denoted by $T_{cp}$, is greater than maximum delay spread of the multipath channels, denoted by $\tau$, i.e., $T_{cp}\gg \tau$. Hence, they require extra overhead to avoid failure in communications. In \cite{rev3}, a noncoherent detection strategy was investigated to solve the noncoherent maximum likelihood (ML) detection problem for a general Q-ary signal constellation which uses a suboptimal energy detector. Furthermore, in \cite{of1,F-Guanding}, an AmBC method was investigated to send data over null (i.e., guard-band) subcarriers of an OFDM signal; however, it does not have a closed-form
expression for the bit detection threshold at the reader. Eventually, the main remarks of the aforementioned existing methods are summarized as follows.

\begin{itemize}
    \item There is a severe spectrum inefficiency in the aforementioned AmBC methods since they either do not consider the physical layer structure of the OFDM signals or send only one bit data over one OFDM symbol regardless of the number of subcarriers. 
    \item Moreover, the communication fails in presence of a severe frequency-selective channel conditions. In other words, they mainly work in a very short-range and poor-scattering environment with a strong line-of-sight (LoS) environments. 
    \item In addition, there is an ignorance of interference caused by AmBC for the legacy receivers due to same spectrum utilization.
\end{itemize}

In this paper, we aim at filling these gaps by proposing a subcarrier-wise backscatter communication over ambient OFDM. In particular, we derive 
a basis transmission scheme and its two  modifications to improve the data rate with a satisfactory BER performance as follows. 
\begin{itemize}
    \item In the basis scheme, it can transmit one bit per subcarrier using on-off keying (OOK) modulation 
    to improve the data rate of the system significantly.
    \item However, since low power IoT devices are mostly accompanied by low/medium data transmission rates (i.e., $<1$ Mbps \cite{AccessSurvey,NB}), in the first modification, referred to as \textbf{Modification-\rom{1}}, one bit is transmitted over an interleaved block of subcarriers; where  repetition coding is used for one bit in each block of subcarriers to improve the BER performance of the system over frequency-selective channels significantly. 
    \item Furthermore, instead of struggling to cancel the direct-link interference at the reader for a better BER performance, like the existing methods in \cite{of-0,of,of1,F-Guanding,Sensor,rev3}, we simply sacrifice a portion of subcarriers to boost the BER performance of our approach profiting from repetition coding.
    \item It is noteworthy that although the superior BER performance is obtained at the expense of data rate (because of using cluster of subcarriers for one bit transmission), \textbf{Modification-\rom{1}} still supports a higher data rate than the existing methods.
    \item Finally, in the second modification, referred to as \textbf{Modification-\rom{2}}, an interleaved index modulation (IM) is employed not only to compensate the decrease of data rate of the \textbf{Modification-\rom{1}}, but also control the power of the signal resulting in interference reduction for a legacy receiver.
\end{itemize}





The rest of the paper is organized as follows. In Section \rom{2}, the system model of subcarrier-wise backscatter communication over ambient OFDM is described. This is the basis scheme for other two modifications. Section \rom{3} provides signal detection and BER analysis of the basis scheme. Then, \textbf{Modification-\rom{1}} and \textbf{Modification-\rom{2}} are proposed and described in Section \rom{4}. Simulation results and comparisons are given in Section \rom{5}. Finally, Section \rom{6} concludes the paper.

 \textit{Notation:} $(.)^T$ and $\lfloor .\rfloor$ denote the transpose operation and the floor operation, respectively. The 2-norm and absolute value of $\mathbf{a}$ and $a$ are denoted by $||\mathbf{a}||$ and $|a|$, respectively. $\mathcal{CN} (\mathbf{a},\mathbf{R})$ represents the distribution of circularly symmetric complex Gaussian (CSCG) random vectors with mean vector $\mathbf{a}$ and covariance matrix $\mathbf{R}$. $\mathbb{C}$ and $\mathbb{R}$ represent sets of complex and real numbers, respectively. 
The Gaussian Q-function is given by $\mathcal{Q}(x)=\frac{1}{\sqrt{2\pi}}\int^\infty_x e^{\frac{-z^2}{2}} dz$.

%% file: Sections/2_system.tex
\label{sec2}
In this section, we describe the system model of our subcarrier-wise backscatter communication over OFDM signals utilizing OOK modulation as shown in Fig. \ref{fig:sys}. This is considered to be a basis scheme for the other two modifications.
We describe our AmBC method in three stages as follows.

\subsection{First Stage}
At the first stage, at the source (e.g., Wi-Fi AP in Fig. \ref{fig:sys}), an OFDM symbol with a
total of $N$ subcarriers is transmitted, which is given by
\begin{equation}
\mathbf{x}=
\mathbf{U}\, \mathbf{s},
 \label{eq11}
\end{equation}
where $\mathbf{s}=[s_1,\, s_2,\, \dots, s_{N}]^T$ is the vector of modulated data symbols. We omit the guard-band subcarriers for simplicity of the notations to avoid any confusion since it does not affect our system. 
In (\ref{eq11}), $\mathbf{U}\triangleq [ \mathbf{M}^T \, \mathbf{F}_{N}^{-1}]^T$ where $\mathbf{F}_{N} \in \mathbb{C}^{N\times N}$ is the discrete Fourier transform (DFT) matrix and $\mathbf{M}$ represents the last $N_{cp}$ rows of the $\mathbf{F}_{N}^{-1}$ for CP \cite{jinB1}.

\subsection{Second Stage}
At the second stage, the OFDM symbol, i.e., $\mathbf{x}$, with length of
$N_s= N+N_{cp}$, reaches to the tag and the reader as shown in Fig.  \ref{fig:sys}. The received ambient OFDM symbol at the reader is given by
\begin{equation}
    \mathbf{y}_a=\mathbf{H}_a\, \mathbf{x},
    \label{eqA1}
\end{equation} 
and at the tag is expressed as
\begin{equation}
    \mathbf{u}=\mathbf{H}_b\, \mathbf{x},
    \label{eqA2}
\end{equation} 
where $\mathbf{H}_a$ and $\mathbf{H}_b$ are the $N_s\times N_s$ lower triangular Toeplitz filtering matrices with the first column of $\mathbf{h}_a=[h_{a,o} , \cdots , h_{a,\ell_a},\, 0, \cdots , 0 ]^T$ and $\mathbf{h}_b=[h_{b,o} , \cdots , h_{b,\ell_b},\, 0, \cdots , 0 ]^T$, respectively.  They
correspond to the channel impulse responses of the direct-link and the forward-link channels from the source to the
reader and the tag, respectively.
Here, $\ell_a$ and $\ell_b$ represent the channels orders and we assume $\mathfrak{L}=\max (\ell_a, \ell_b) \leqslant N_{cp}$, i.e., $\tau \leqslant T_{cp}$.
\begin{figure}[t]
\centering
\captionsetup{width=0.8\linewidth}
\subfloat[The tag's RF-analog front-end. Power of the ambient RF signal is transferred from the antenna to the rectifier through the matching network. The rectifier supplies the needed DC power. Once the tag is powered up\protect\footnotemark, $\mathbf{u}$ is being filtered and transmitted.]{%
 \includegraphics[width=5.8cm, height=2.7cm]{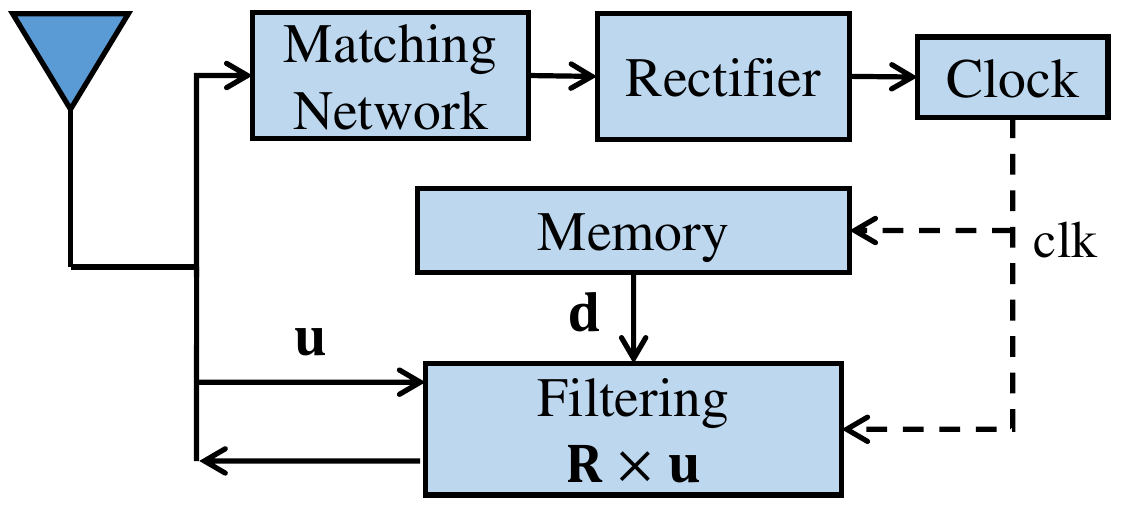}%
 } 
 \vspace{0.01\linewidth}
\captionsetup{width=.8\linewidth}
\subfloat[Passive notch filter bank is used to pass or halt different OFDM subcarriers with regard to $\mathbf{d}$.]{%
 \includegraphics[width=6.6cm, height=3.5cm]{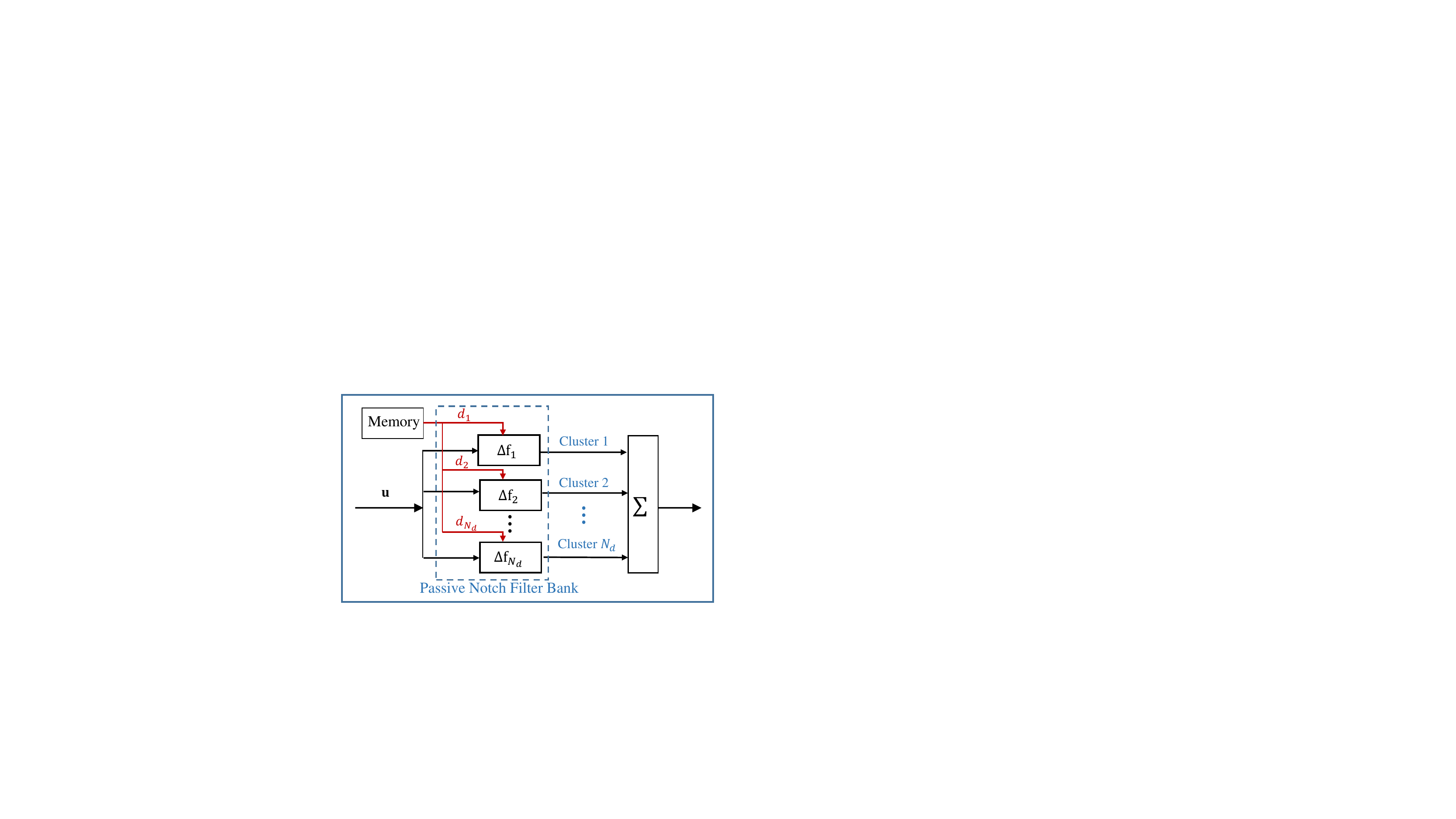}%
  \label{fig:tagB}}
 \captionsetup{width=1\linewidth}
 \caption{ Tag receives the ambient OFDM signal and modulates its subcarriers
with OOK modulation. $l^{\text{th}}$ filter nullifies $l^{\text{
th}}$ cluster of subcarriers when $d_l = 0$ and
passes it when $d_l = 1$.} 
 \label{fig:tag}
\end{figure}
\footnotetext{It is noteworthy that the harvested power is a nonlinear function of the input RF power \cite{rev3H1,H34}. Moreover, when the input power is below the harvesting circuit power sensitivity threshold, the harvested power is zero \cite{revH2}. In this study, we assume that the harvested power is enough to trigger the tag. Nevertheless, our proposed method is not limited to the passive tags and a secondary battery can be used like in semi-passive tags if needed.}

At the tag, the desired data is a vector $\mathbf{d}=[d_1, \, d_2,\, \dots, \, d_{N_d}]^T$, where $d_l\,( l=1,...,\, N_d) $ represents the data symbol. 
If OOK is employed, we can have $d_l \in \{0, 1\}$. Throughout the paper, we assume that $d_l$ is regarded as a bit
for convenience.
Fig. \ref{fig:tag} illustrates the tag's RF-analog front-end which has a filter bank of passive notch filters. We consider a matrix,
denoted by $\mathbf{R}$, that works as a filtering matrix to pass or nullify the $l^{\text{th}}$ cluster of subcarriers corresponding to $d_l = 1$ and $d_l=0$, respectively. Passive analog notch filters can be employed for this
task. 

\subsubsection{Discussion on Feasibility of Filtering}
\begin{figure}[t]
\centering
\captionsetup{width=1\linewidth}
  \includegraphics[width=7.5cm, height=5cm]{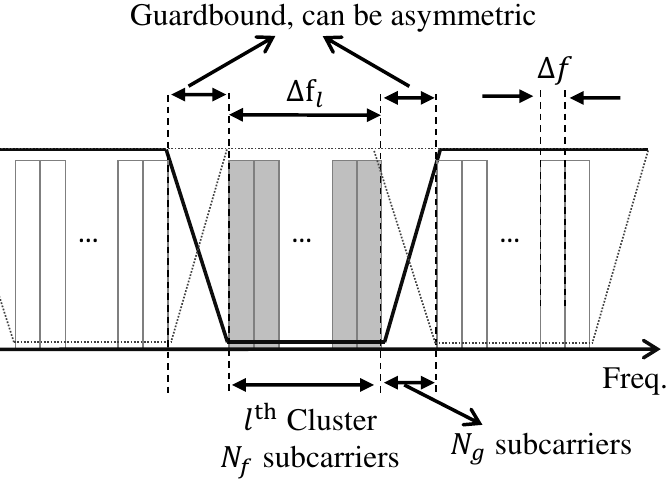}%
  \caption{Illustration of RF notch filter over OFDM subcarriers. The $l^{\text{th}}$ RF notch filter with bandwidth of $\Delta \mathsf{f}_l$ is controlled by $d_l$ to pass or halt the $l^{\text{th}}$ cluster of subcarriers.}
  \label{fig:clus}
\end{figure}

In order to put the filtering matrix $\mathbf{R}$ into action, 
we need to have highly selective filters due to a narrow subcarrier spacing of $\Delta f$ typically in UHF\footnote{UHF: Ultra High Frequency} band.
Let $\Delta \mathsf{f}_l$ and $f_c$ denote the bandwidth and center frequency of the $l^{\text{th}}$ notch filter in the filter bank $\mathbf{R}$, respectively (as shown in Fig. \ref{fig:tag} (b)). It is ideal to have $\Delta \mathsf{f}_l=\Delta f$. However, finding perfect high-order passive RF filters with $\Delta \mathsf{f}_l=\Delta f$ in UHF band, e.g., $f_c=2.4$ GHz in IEEE 802.11, is challenging. 
We still do not find such an appropriate passive RF notch filter as of writing of this paper. 
Nevertheless, this challenge is similar to the traditional and well-known issue of frequency selecting/planning in cognitive radio transceivers where the scarce spectrum, e.g., UHF spectrum, is utilized densely \cite{SurFil,Theory}. 
Standard reflective and absorptive filters are suitable for most of these applications; however, absorption filters are better choices in performance as they provide greater levels of attenuation and frequency selectivity besides their non-reflective properties \cite{ Theory,SurFil,  Awli,Awli2,        Cascaded,   FreqAgile2,FreqAgile,FSEL}. 
The passive RF notch filters in \cite{Awli,Awli2,Cascaded,Theory} provide a stopband with range of $\Delta \mathsf{f}_l\approx 0.5-4$ MHz in UHF band.
For instance, the passive notch filter in \cite{Awli} provides a highly selective narrow rejection with a narrow $10$ dB rejection stopband of $\Delta \mathsf{f}_l=2.4$ MHz over $f_c=3.4$ GHz. Moreover, the notch filters in \cite{Theory,Awli2} can provide a higher frequency selectivity with a fractional bandwidth of up to $0.02\%$ in UHF band, i.e.,  $\Delta \mathsf{f}_l\approx 480$ kHz over $f_c=2.4$ GHz. Recently, in \cite{TranFil,SurFil}, RF notch fitters are proposed with even more tunable bandwidth from $\Delta \mathsf{f}_l=0$ MHz up to $\Delta \mathsf{f}_l=96$ MHz over higher frequencies up to $f_c=3.4$ GHz. %
Besides, these RF notch filters have steep roll-off responses in the transition band between the stopband and passband.
However, considering the current OFDM standards, e.g., IEEE 810.11a, the stopband of the RF notch filters is still mostly much wider than one OFDM subcarrier spacing, e.g., $\Delta f=312.5$ kHz, i.e., $\Delta \mathsf{f}_l > \Delta f$. 
Therefore, we utilize the feasibility of aforementioned RF notch filter concepts and consider that the stopband of notch filters in the filter bank covers a cluster of $N_f$ subcarriers with $N_g$ subcarriers as its guardband shown in Fig. \ref{fig:clus}.
So, the total number of required subcarriers becomes
\begin{equation}
    N=N_d (N_f)+(N_d-1)N_g.
    \label{nnD}
\end{equation}
 In \eqref{nnD}, let consider $N$ is fixed, then the total number of data symbols, i.e., $N_d$, is given by
\begin{equation}
    N_d=\frac{N+N_g}{N_f+N_g}.
\end{equation}
It is evident that a larger $\Delta \mathsf{f}_l$ is easier for filter bank implementation at the cost of data rate.
It is also noteworthy that using a large number of passive RF components (e.g., different load impedance states) is a routine technique to reach higher modulation orders in the conventional AmBC models \cite{LISA,BPfilter}. Accordingly, in our model, various passive RF components can be used for filtering different
subcarrier frequencies.
Note that thermal noise is neglected in the incident signal at the tag because it does not have any active RF component, which is a common assumption for passive tags and is supported in \cite{of1,of,of-0,F-Guanding,Sensor} as well. However, the signal is attenuated by a
factor of $\beta$ inside the tag. 
There is extensive efforts in the
literature (e.g., in \cite{rev,Bl-1,Bl-2,Bl-3,BPfilter,rev3H1,revH2} and references therein) where similar
passive tag structures with passive filters are proposed. 
However, evaluating the electronic circuit of the tag in
details is beyond the scope and space of this paper.

{\it Assumption 1 (Highly Selective Notch Filter):} In the analysis given in the next parts, we assume to have near-perfect high-quality factor filters similar to the notch filters in \cite{Theory,Awli2, TranFil} that are able to satisfy the condition of 
\begin{equation}
    \Delta \mathsf{f}_l \leqslant \Delta f \rightarrow N_f=1,\, N_g=0,
    \label{assu1}
\end{equation}
 in order to simplify the analysis and verify the feasibility of the preliminary
idea. A similar assumption has been taken into account in \cite{wifiOFDMA}. 

\subsubsection{Backscattered Signal}
Eventually, the tag backscatters the filtered signal to the reader. The
backward-link channel from the tag to the reader has a single path \cite{of,of1,chan2} because the distance between the tag and the reader has to be sufficiently close in backscatter communication; and its channel coefficient is denoted by $h_c$ as
shown in Fig. \ref{fig:sys}. The received backscattered signal at the reader
is given by
\begin{align}
\mathbf{y}_b&= h_c\,\beta\, \mathbf{R}\, \mathbf{u}.
\label{eq4}
\end{align}

\subsection{Third Stage}
Finally, at the third stage, at the reader, the received signal with the background noise  $\mathbf{n}\sim \mathcal{CN} (0, N_0)$ is given by
\begin{equation}
\mathbf{y}=\mathbf{y}_a+\mathbf{y}_b+\mathbf{n}.
\label{eqs1}
\end{equation}
It is worth mentioning that the length of the OFDM symbol is much longer than the propagation delay due to short distance between the tag and the reader, and we ignore the propagation delay \cite{of}. 

At the reader side, the CP portion is discarded and the remaining part, denoted by vector $\mathbf{y}_d$, goes through the DFT operation. 
 
\begin{align}
    \mathbf{r}&= \mathbf{F}_N \mathbf{y}_d \nonumber\\
    &=
    \mathbf{F}_N \mathbf{H}^c_a \mathbf{F}_N^{-1} \mathbf{s}+ 
     \mathbf{F}_N  h_c\,\beta\, \mathbf{R}^c\, \mathbf{H}^c_b \mathbf{F}_N^{-1} \mathbf{s} + \tilde{\mathbf{n}},
     \label{rtg}
\end{align}
where $(.)^c$ denotes the circularized matrix reflecting the effect of CP after CP removal at the receiver; and $\mathbf{\tilde{n}}$ is the background noise in the frequency domain. 

Since filtering matrix $\mathbf{R}^c$ acts as a frequency selective matrix to nullify or preserve the subcarriers in the frequency domain with respect to $\mathbf{d}$, and also with respect to the {\it Assumption 1}, its response in the discrete time domain can notationally be defined as $\mathbf{R}^c\triangleq \mathbf{F}^{-1}_N \left[\text{diag}(\mathbf{d})\right] \mathbf{F}_N$. Thus, from \eqref{rtg}, we have
\begingroup\makeatletter\def\f@size{9.5}\check@mathfonts
\begin{equation}
    \mathbf{r}= 
    \overbrace{\mathbf{F}_N \mathbf{H}^c_a \mathbf{F}_N^{-1}}^{\substack{\text{Frequency response}\\\text{of }\mathbf{h}_a}} \mathbf{s}+ 
      h_c\,\beta\, \overbrace{\mathbf{F}_N \mathbf{F}^{-1}_N}^{\mathbf{I}_N} \left[\text{diag}(\mathbf{d})\right] \overbrace{\mathbf{F}_N \mathbf{H}^c_b \mathbf{F}_N^{-1}}^{\substack{\text{Frequency response}\\\text{of }\mathbf{h}_b}}  \mathbf{s} + \tilde{\mathbf{n}}.
\end{equation}
\endgroup

Therefore, the received vector in the frequency domain becomes

\begin{equation}
    \mathbf{r}= \mathbf{\tilde{H}}_a \mathbf{s} +  h_c \,\beta\, \mathbf{\tilde{H}}_b\, \left[\text{diag}(\mathbf{d})\right]\, \mathbf{s} + \mathbf{\tilde{n}},
    \label{eqs}
\end{equation}
where $\mathbf{\tilde{H}}_a$ and $\mathbf{\tilde{H}}_b$ are the $N\times N$ diagonal matrices with $\mathbf{\tilde{h}}_a=\mathbf{F}_{N}\mathbf{h}_a=[\tilde{h}_{a,0},\cdots , \tilde{h}_{a,N}]^T$ and $\mathbf{\tilde{h}}_b=\mathbf{F}_{N}\mathbf{h}_b=[\tilde{h}_{b,0},\cdots , \tilde{h}_{b,N}]^T$ for their diagonal elements, respectively.

The numbers of active and inactive subcarriers in the backscattered signal, i.e., $\mathbf{y}_b$ in (\ref{eq4}), are not fixed as they depend on $\mathbf{d}$ which is a random data vector. To be more specific, the numbers of active and inactive subcarriers equal $Q=||\mathbf{d}||_0$ and $K=N-Q$, respectively. Let define two supports for $\mathbf{d}$ as $I_{1}\triangleq \{l|[\mathbf{d}]_l=1\}$ and $I_{0}\triangleq \{l|[\mathbf{d}]_l=0\}$. Then, the total signal-to-noise-ratio (SNR) of the received signal in (\ref{eqs}) is defined as the total received signal power, i.e., including both direct-link and bachscattered signals, at the reader over the background noise power; and can be expressed as
\begin{equation}
    \Gamma= \frac{ \sum\limits_{l\in I_0} |\tilde{h}_{a,l}|^2 +  \, \sum\limits_{l\in I_1} |\tilde{h}_{a,l}+h_c\beta\tilde{h}_{b,l}|^2 }{N}\times\frac{E_b}{N_0},
\end{equation}
where $E_b$ stands for bit energy.
Moreover, from (\ref{eqs}), the received signal through the $l^{th}$ subcarrier, denoted by $r_l$, is 
\be
r_l = (\tilde{h}_{a,l} + \overbrace{h_c\beta \tilde{h}_{b,l}}^{\tilde{h}_{s,l}} d_l) s_l + \tilde{n}_l,\quad \ l = 1,\ldots, N_s,
	\label{EQ:r2}
\ee
where $\tilde{h}_{s,l}\in \mathbf{\tilde{h}}_s\left(=h_c\beta\mathbf{\tilde{h}}_{b}\right)$ denotes the composite source-tag-reader channel coefficient as shown in Fig. \ref{fig:sys} which is usually referred to as the dyadic-backscatter-link in the literature \cite{LISA,dya}. Consequently, the SNR of the received signal over the $l^{\text{th}}$ subcarrier is given by
\begin{equation}
    \left \{ \begin{array}{l}
         \Gamma_l\bigg|_{d_l=0} =  \frac{E_b |\tilde{h}_{a,l}|^2}{N_0}, \vspace{1mm}\\
         \Gamma_l\bigg|_{d_l=1} =  \frac{E_b |\tilde{h}_{a,l}+\tilde{h}_{s,l}|^2}{N_0}  .
    \end{array} \right.   
\end{equation}



\subsection{Data Transmission Rate Analysis}
While the existing methods in \cite{of-3,Jin_OF,of-0,of,Sensor} send one bit over one OFDM symbol period, our proposed method can transmit up to $N_d>1$ bits over one OFDM symbol period. For example, consider an OFDM system based on IEEE 802.11a specifications (i.e., OFDM symbol duration of $T=4\,\mu$sec, DFT size of $64$, $\Delta f=312.5$ KHz, and $N=52$ used subcarriers), then, the existing methods can support a data rate of up to $\eta=\frac{1}{4\,\mu\text{sec}}=250$ kbps; while our proposed method can support a data rate of $\eta=\frac{N_d}{4\,\mu\text{sec}}$. If we assume $\Delta \mathsf{f}_l$ varies between $0.5-4$ MHz based on the current filters in \cite{TranFil,Awli2,Theory,Awli}, i.e., $N_f\approx 1$ to $12$ subcarriers, then, $N_d\approx 52$ to $4$ subcarriers and we can expect a variable data rate from $1$ MHz up to $13$ MHz.
Such a data rate performance is up to $N_d$ times better than the existing AmBC systems in the literature.

%% file: Sections/3_detection.tex
\label{sec3}
 We consider the reader is able to detect the pilots
\footnote{Similar to the fixed pilot pattern in LTE standard for slow fading channel estimation, the downlink pilots are unique complex numbers dependent on the source and subcarriers \cite{nutshel}.} from the source. Then, the reader can estimate the both the direct-link and dyadic-backscatter-link CSIs. In other words, they can be estimated by letting the source send pilots reaching the reader through the direct-link $\mathbf{\tilde{h}}_{a}$ and dyadic-backscatter-link $\mathbf{\tilde{h}}_{s}$ channels. Since there are two different paths, two estimations are required. 
First, the tag initiates a zero data vector, i.e., $||\mathbf{d}||_0=0$, and from \eqref{eqs}, the reader receives $\mathbf{r}=\tilde{\mathbf{H}}_a \mathbf{x}_p+\mathbf{n}$ where $\mathbf{x}_p$ is the pilot OFDM symbol. Subsequently, reader estimates $\mathbf{\tilde{h}}_{a}$. Second, the tag initiates $\forall l\rightarrow d_l=1$, i.e., $||\mathbf{d}||_0=N_d$, and from \eqref{eqs}, the reader receives $\mathbf{r}=\left(\tilde{\mathbf{H}}_a + \tilde{\mathbf{H}}_s\right) \mathbf{x}_p+\mathbf{n}$ and $\mathbf{\tilde{h}}_{s}$ is estimated this time\footnote {$\tilde{\mathbf{H}}_s$ is $N\times N$ diagonal matrix with $\tilde{\mathbf{h}}_s$ for its diagonal elements.}.
 Therefore, $\mathbf{\tilde{h}}_{a}$ and $\mathbf{\tilde{{h}}}_s$ are assumed to be estimated by minimum mean square error (MMSE) channel estimation technique \cite{jinB1,bookmimo} at the reader, which is similar to the assumption in \cite{rev} for coherent signal detection.
 
{\it Assumption 2 (Near-Perfect CSI Estimation):} We assume near-perfect CSI estimation at the reader, i.e., the direct-link $\mathbf{\tilde{h}}_{a}$ and dyadic-backscatter-link $\mathbf{\tilde{h}}_{s}$ channels are estimated and known to the reader.

\begin{table*}[bp]
\def\svgwidth{\textwidth}
\noindent\rule{\textwidth}{1pt}
\begin{align}
    e^{- \frac{1}{N_0} |r_l - \tilde{h}_{a,l} \sqrt{E_b}|^2}+
    e^{- \frac{1}{N_0} |r_l + \tilde{h}_{a,l} \sqrt{E_b}|^2}
    \defh
    e^{- \frac{1}{N_0} |r_l - (\tilde{h}_{a,l}+\tilde{h}_{s,l}) \sqrt{E_b}|^2}+
    e^{- \frac{1}{N_0} |r_l + (\tilde{h}_{a,l}+\tilde{h}_{s,l}) \sqrt{E_b}|^2},\tag{19}\label{er43}
\end{align}\vspace{-10mm}\\
\begin{align}
    e^{-\frac{1}{N_0}\tilde{h}_{a,l}^2 E_b}\overbrace{\left[
    e^{+ \frac{1}{N_0} 2 r_l \tilde{h}_{a,l}\sqrt{E_b}}+
    e^{-\frac{1}{N_0} 2 r_l \tilde{h}_{a,l}\sqrt{E_b}}\right]}^
    {2\cosh\left(\frac{2 r_l \tilde{h}_{a,l}\sqrt{E_b}}{N_0} \right)}
    \defh
    e^{-\frac{1}{N_0}(\tilde{h}_{a,l}+\tilde{h}_{s,l})^2 E_b}\overbrace{\left[
    e^{+ \frac{1}{N_0} 2 r_l (\tilde{h}_{a,l}+\tilde{h}_{s,l})\sqrt{E_b}}+
    e^{-\frac{1}{N_0} 2 r_l (\tilde{h}_{a,l}+\tilde{h}_{s,l})\sqrt{E_b}}\right]}^
    {2\cosh \left(\frac{1}{N_0} 2 r_l (\tilde{h}_{a,l}+\tilde{h}_{s,l})\sqrt{E_b}\right)}.\tag{20}\label{rte2}
\end{align}
\end{table*}
\subsection{Signal Detection}\label{2.A.2}
In order to detect the signal, an energy detector over the subcarriers is proposed. It is noteworthy that the proposed detection method does not require direct-link interference cancellation. In the following, an optimal decision threshold for the received power and the test statistic are derived.
Let $\beta=1$ for notational simplicity. Then, at the reader, the conditional distribution of $r_l$ on $s_l$ and $d_l$
is given by
\be
f(r_l\,|\, d_l, s_l) 
= \frac{1}{\sqrt{\pi N_0}} \exp
\left(
-\frac{1}{N_0}| r_l - (\tilde{h}_{a,l} + \tilde{h}_{s,l} d_l) s_l|^2
\right).
\ee
Considering that $d_l$ is equally likely
(i.e., $\Pr(d_l = 0) = \Pr(d_l = 1) = \frac{1}{2}$),
the maximum likelihood (ML) detection is equivalent to the maximum posteriori probability (MAP) detection \cite{jinB2}, which is as follows
\be
f(r_l \,|\, d_l = 0, s_l) 
\defh f(r_l \,|\, d_l = 1, s_l),
\ee
where $H_i$, $i\in\{0,\,1\}$, represents that $d_l=i$.
 Here, $s_l\in \cS$ is unknown and equally likely (e.g., binary phase shift keying (BPSK)), where $\cS=\{\pm \sqrt{E_b} \}$. Thus, the likelihood function of $d_l$ for given $r_l$
becomes
\be
f(r_l\,|\, d_l) = \sum_{s_l \in \cS} f(r_l\,|\, s_l, d_l) \Pr(s_l).
\ee
The ML detection becomes
\be
\sum_{s_l \in  \cS} 
e^{- \frac{1}{N_0} |r_l - \tilde{h}_{a,l} s_l|^2} \defh
\sum_{s_l \in  \cS} 
e^{- \frac{1}{N_0} |r_l - (\tilde{h}_{a,l}+\tilde{h}_{s,l}) s_l|^2}.
	\label{EQ:Etest}
\ee

Note that this method is also applicable to other modulated symbols with only one level of power, i.e., $E_b$. 
The ML detection in \eqref{EQ:Etest}, after some manipulations given in \eqref{er43} and \eqref{rte2} at the bottom of this page, becomes

\setcounter{equation}{20}
\be
\frac{\cosh \left( \frac{2 \tilde{h}_{a,l} \sqrt{E_b} r_l} {N_0} \right) }
{\cosh \left( \frac{2 (\tilde{h}_{a,l} + \tilde{h}_{s,l}) \sqrt{E_b} r_l} {N_0} \right) }
\defh \exp
\left( - \frac{(2 \tilde{h}_{a,l} \tilde{h}_{s,l} + \tilde{h}_{s,l}^2) E_b}{N_0}
\right).
\ee
At a high SNR, since
\be
\frac{\cosh \left( \frac{2 \tilde{h}_{a,l} \sqrt{E_b} r_l} {N_0} \right) }
{\cosh \left( \frac{2 (\tilde{h}_{a,l} + \tilde{h}_{s,l}) \sqrt{E_b} r_l} {N_0} \right) }
\approx
\exp \left( - \frac{2 \tilde{h}_{s,l} \sqrt{E_b} |r_l|}{N_0} \right),
\ee
the optimal decision threshold, denoted by $\delta_l$, becomes
\be
\delta_l = \frac{(|\tilde{h}_{a,l}| + |\tilde{h}_{a,l}+\tilde{h}_{s,l}|) \sqrt{E_b}}{2}.
\label{ep}
\ee
Eventually, the test statistic can be given as
\begin{subnumcases}{}
|r_l| \underset{H_0}{\overset{H_1}{\gtrless}}  \delta_l  \qquad   \text{when} \quad |\tilde{h}_{a,l}|< |\tilde{h}_{a,l}+\tilde{h}_{s,l}|\label{32er} \\
|r_l| \underset{H_0}{\overset{H_1}{\lessgtr}}  \delta_l  \qquad   \text{when} \quad |\tilde{h}_{a,l}|> |\tilde{h}_{a,l}+\tilde{h}_{s,l}|
\label{ep21}
\end{subnumcases}
We note that in general, the phase information of the channel gains along with their amplitudes is in fact relevant to the detection. This is because with the fading assumption, the channel gains $\tilde{h}_{a,l}$ and $\tilde{h}_{s,l}$ could be added constructively or destructively and therefore both \eqref{32er} or \eqref{ep21} are likely to happen. However, we only need to know whether the sum of these gains are constructive or destructive and to this end, an accurate knowledge of channel phases are not required. In other words, only an estimate of the channel phase information would suffice to determine whether the sum is constructive or destructive.

In the following subsections, the BER performance of the proposed scheme in additive white Gaussian noise (AWGN) channels and frequency-selective channels is studied.
\subsection{BER Analysis in AWGN Channels}\label{sec2:sub1}
 In this subsection, we omit the subcarrier index $l$ for convenience and assume $\beta=1$. Furthermore, assume that the forward-link and direct-link channels can be modeled as AWGN channels thanks to short distances between the source, the reader and the tag \cite{wcnc}. Equation (\ref{EQ:r2}) in AWGN channels is rewritten as
%
\begin{figure}[t]
\centering
\captionsetup{width=1\linewidth}
  \includegraphics[width=6.5cm, height=3.3cm]{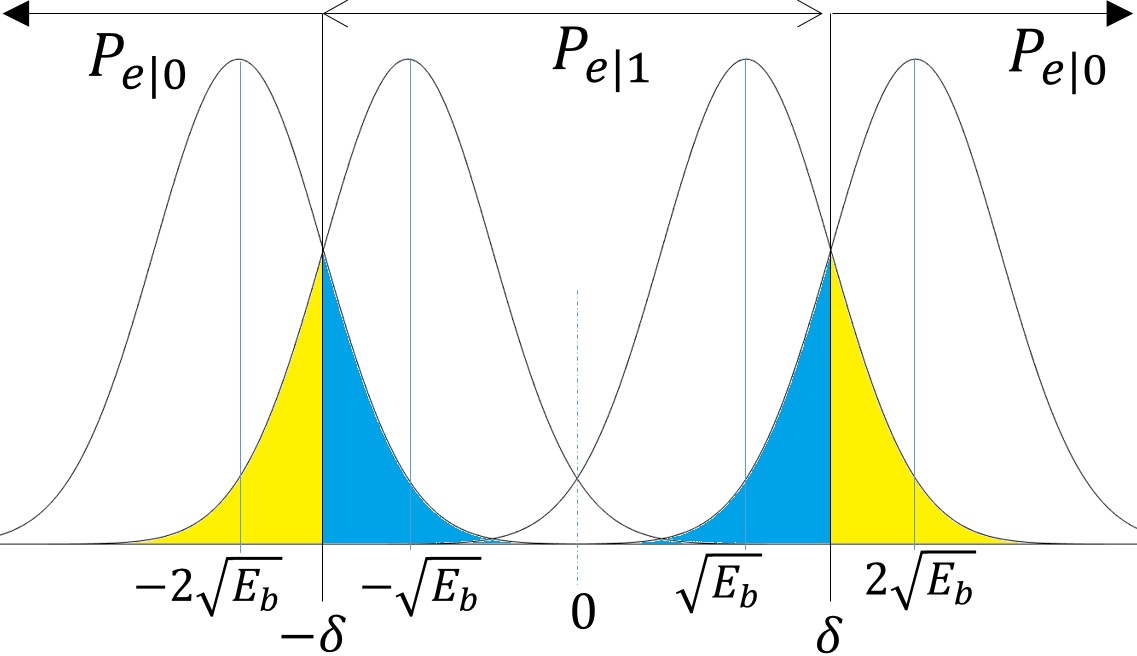}%
  \caption{The likelihood functions of the proposed method and BER regions for $P_{e|0}$ and $P_{e|1}$ in AWGN channels.}%
  \label{fig:pdf}
\end{figure}
\begin{equation}
    r=(1+d)s+\tilde{n}.
\end{equation}
In such a scenario, the reader can see four different subcarrier
amplitudes from the received signal. Fig. \ref{fig:pdf} shows the likelihood functions of possible amplitudes of each subcarrier in
the frequency domain, where the shaded areas represent error
probabilities. Therefore, the error probability when bit 0 is
transmitted is shown in yellow shadowed part of Fig. \ref{fig:pdf} and is given by
\begin{align}
&P_{e|0}  = \nonumber\\ &\frac{1}{2\sqrt{\pi N_0}}\biggl[ \int_{-\infty}^{-\delta} e^{-\frac{(x+\sqrt{E_b})^2}{N_0}} dx + 
                                                                      \int^{\infty}_\delta e^{-\frac{(x+\sqrt{E_b})^2}{N_0}} dx  \nonumber\\
              &  +                                                  \int_{-\infty}^{-\delta} e^{-\frac{(x-\sqrt{E_b})^2}{N_0}} dx+
                                                                      \int^{\infty}_\delta e^{-\frac{(x-\sqrt{E_b})^2}{N_0}} dx \biggr].
\label{pe0}
\end{align}
Substituting $\delta$ from (\ref{ep}) into (\ref{pe0}) with respect to the AWGN channel coefficients, i.e., $\delta=\frac{3\sqrt{E_b}}{2}$, we have
\begin{align}
P_{e|0} = \mathcal{Q}\biggl(\sqrt{\frac{1}{2}\gamma}\biggr) + \mathcal{Q}\biggl(\sqrt{\frac{25}{2} \gamma}\biggr),
\end{align}
where $\gamma\triangleq\frac{E_b}{N_0}$. Likewise, when bit 1 is transmitted the error probability is shown in blue shadowed part of Fig. \ref{fig:pdf} and can be obtained as
\begin{align}
P_{e|1}=1-\mathcal{Q}\biggl(-\sqrt{\frac{1}{2}\gamma}\biggr)-\mathcal{Q}\biggl(\sqrt{\frac{49}{2}\gamma}\biggr).
\end{align}
Finally, the BER in an AWGN channel is given by
\begin{align}
P_{e-\text{AWGN}} &=\mathcal{Q}\biggl(\sqrt{\frac{1}{2}\gamma}\biggr)+\frac{1}{2}\mathcal{Q}\biggl(\sqrt{\frac{25}{2}\gamma}\biggr)- \frac{1}{2}\mathcal{Q}\biggl(\sqrt{\frac{49}{2}\gamma}\biggr) \nonumber\\ 
& \approx \mathcal{Q}\biggl( \sqrt{\frac{1}{2}\gamma}\biggr).
\end{align}

Next, the BER performance of the proposed model in frequency-selective channels is studied.

\subsection{BER Analysis in Frequency-Selective Fading Channels}\label{SBER} 
\begin{figure}[t]
\centering
\captionsetup{width=1\linewidth}
  \includegraphics[width=7cm, height=4.9cm]{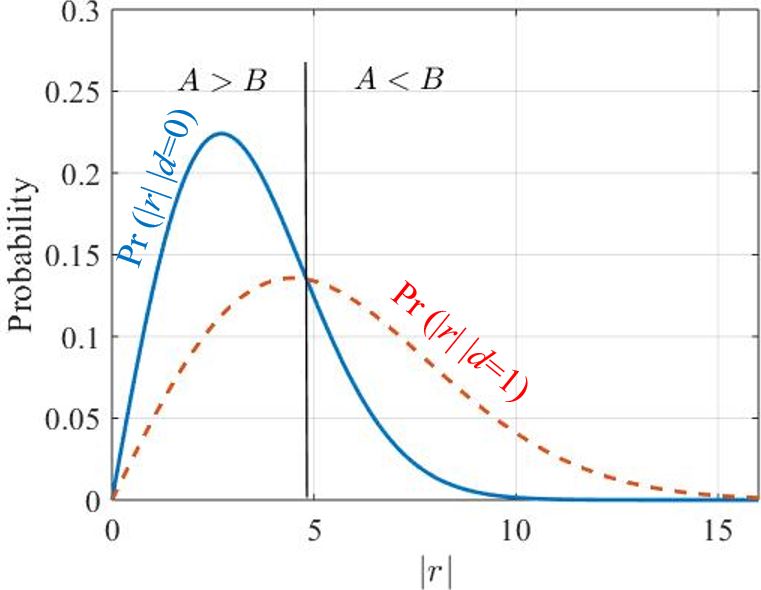}%
  \caption{Probability distribution functions of $|r|$ when $d=0$ and $d=1$. $\gamma=5$dB, $\sigma_a^2=2$, $\sigma_b^2=2$, $\sigma_c^2=2$.}%
  \label{fig:pr}
\end{figure}
In this subsection, we consider the case that the source is not sufficiently close to the reader and the tag. As a result, the forward-link and direct-link channels can be modeled as frequency-selective fading channels.
Suppose $d=0$ is transmitted. From (\ref{EQ:r2}), we have
\begin{equation}
r\bigg|_{d=0}=\tilde{h}_{a} s +\tilde{n}. \hspace{1.2cm}
\end{equation}
Likewise, if $d=1$ is transmitted, the received signal is
\begin{equation}
r\bigg|_{d=1}=(\tilde{h}_a+\tilde{h}_s) s +\tilde{n}.
\end{equation}
Throughout the paper, for Rayleigh fading, we consider $\tilde{h}_a \sim \mathcal{CN}(0,\sigma_a^2)$, $\tilde{h}_b \sim \mathcal{CN}(0,\sigma_b^2)$ , $\tilde{h}_c \sim \mathcal{CN}(0,\sigma_c^2)$ which are common assumptions in the literature, e.g., \cite{choi2017,GFIM}. Suppose that $\tilde{h}_b$ and $\tilde{h}_c$ are independent, then we have $\tilde{h}_s \sim \mathcal{CN}(0,\beta^2\sigma_b^2\sigma_c^2)$. We consider $h_a$ and $h_s$ are also independent. Since $(\tilde{h}_a+\tilde{h}_s)\sim \mathcal{CN}(0,\sigma_a^2+\beta^2\sigma_b^2\sigma_c^2)$, we have
\begin{align}\left\{
\begin{array}{l}
     r\big|_{d=0}\sim \mathcal{CN}(0,E_b\sigma_a^2+ N_0), \vspace{1mm}\\
     r\big|_{d=1}\sim \mathcal{CN}(0,E_b(\sigma_a^2+\beta^2\sigma_b^2\sigma_c^2)+ N_0).
\end{array}\right.\label{eqte}
\end{align}

The absolute values of $r$ in (\ref{eqte}) have Rayleigh distributions \cite{choi2017,GFIM}. The probability distribution functions (PDFs) of them are shown in Fig. \ref{fig:pr}, where $A\triangleq|\tilde{h}_a|$ and $B\triangleq|\tilde{h}_a+\tilde{h}_s|$ for notational simplicity.

\begin{figure}[t]
\centering
\captionsetup{width=1\linewidth}
\subfloat[The likelihood functions if $A< B$]{%
 \includegraphics[width=6.5cm, height=3.3cm]{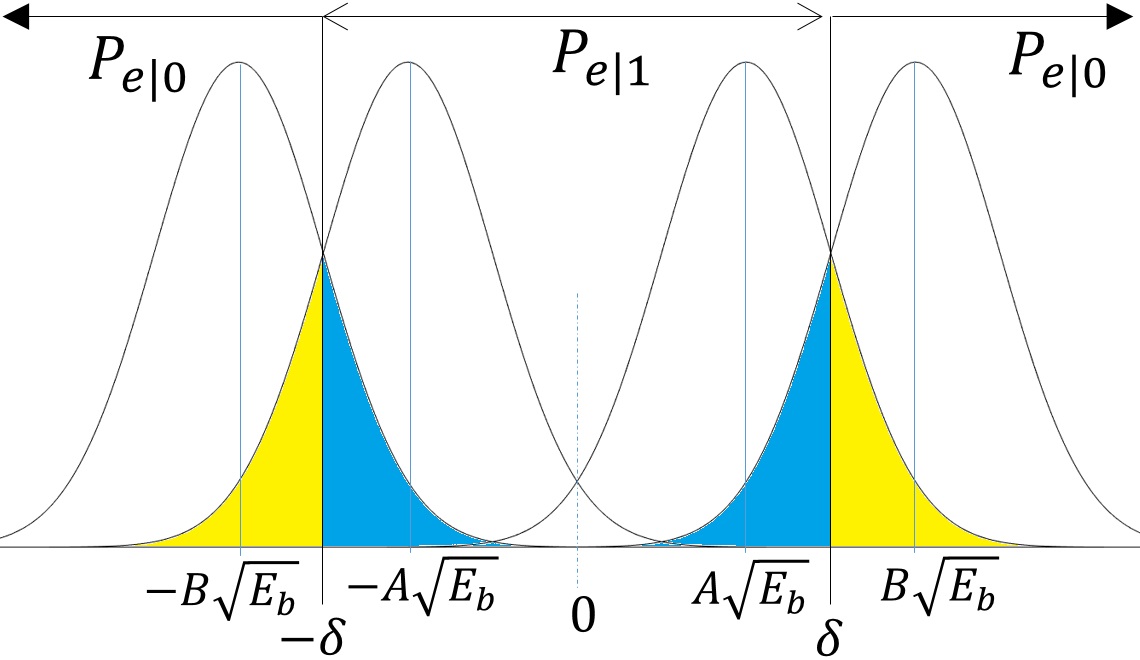}%
 } 
 \vspace{0.01\linewidth}
\captionsetup{width=1\linewidth}
\subfloat[The likelihood functions if $A>B$]{%
 \includegraphics[width=6.5cm, height=3.3cm]{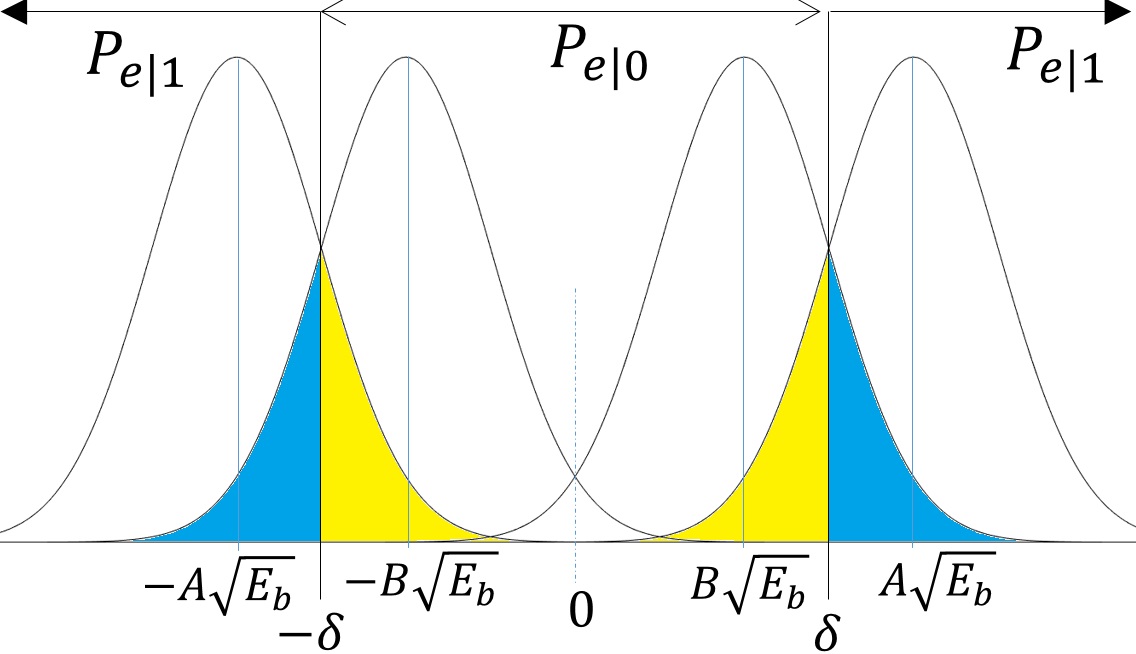}%
  \label{fig:tagB}}
 \captionsetup{width=1\linewidth}
 \caption{ The likelihood functions and BER regions for $P_{e|0}$ and $P_{e|1}$ for two different possible scenarios in frequency-selective channels.}
 \label{fig:LIFreq}
\end{figure}

As a result, the test statistic is re-written for two different possible scenarios in frequency-selective channels as
\begin{equation}
\left \{ \begin{array}{c}
|r| \underset{H_0}{\overset{H_1}{\gtrless}}  \delta  \qquad   if \quad A< B \quad \text{, Fig. \ref{fig:LIFreq} (a)}\vspace{2mm}\\
|r| \underset{H_0}{\overset{H_1}{\lessgtr}}  \delta  \qquad   if \quad A>B  \quad \text{, Fig. \ref{fig:LIFreq} (b)}
\end{array} \right. ,
\label{eqdeli}
\end{equation}
where $\delta =\frac{A+B}{2}\sqrt{E_b}$ is derived from (\ref{ep}).

Suppose that $A<B$ as shown in Fig. \ref{fig:LIFreq} (a), then, after analysis given in Appendix \ref{AP1}, the following probabilities are given.
\begin{equation} 
  \left\{\begin{array}{c}
       \Pr(|r|>\delta \, \big|d=0,A<B,B)\approx\mathcal{Q}\left ((B-A)\sqrt{\frac{\gamma}{2}} \right) \vspace{1mm} \\
       \Pr(|r|<\delta \, \big|d=1,A<B,B)\approx\mathcal{Q}\left ((B-A)\sqrt{\frac{\gamma}{2}} \right) 
  \end{array} \right. . 
  \label{eqAP1}
\end{equation}
Consequently, we have 
\begin{equation}
    P_e(|r|\big|A<B,B)= \mathcal{Q}\left ((B-A)\sqrt{\frac{\gamma}{2}} \right).
    \label{eq_pe1}
\end{equation}
Likewise, when $A>B$,
\begin{equation}
 \left\{\begin{array}{c}
     P(|r|<\delta \,\big|d=0,A>B,B)\approx\mathcal{Q}\left ((A-B)\sqrt{\frac{\gamma}{2}} \right)\vspace{1mm}\\
     P(|r|>\delta\, \big|d=1,A>B,B)\approx\mathcal{Q}\left ((A-B)\sqrt{\frac{\gamma}{2}} \right)
       \end{array} \right. .
       \label{eqap12}
\end{equation}
Then, similar to (\ref{eq_pe1}), we have
\begin{equation}
     P_e(|r|\,\big|A>B,B)= \mathcal{Q}\left ((A-B)\sqrt{\frac{\gamma}{2}} \right) .
     \label{eq_pe2}
\end{equation}
Suppose that $A$ and $B$ are Rayleigh random variables with the PDF of $f(x)=\frac{x}{\Omega}\exp (-\frac{x^2}{2\Omega})$, where $\Omega$ is the average power. Then, the error probability of the basis scheme can be obtained from (\ref{eq_pe1}) and (\ref{eq_pe2}) as 
\begingroup\makeatletter\def\f@size{9}\check@mathfonts
\begin{align}
    P_{e|\textbf{basis}}=&\int_0^\infty  \left[\int_0^B P_e(|r|\big|A<B,B) f(A) \, dA\right] f(B)\, dB +\nonumber \\
    & \int_0^\infty \left[\int_B^\infty P_e(|r|\big|A>B,B) f(A) \, dA \right] f(B) \, dB.
    \label{pEE}
\end{align}  
\endgroup
Let $\mu\triangleq \sqrt{\frac{E_b}{N_0}}$, $\alpha \triangleq \frac{\mu^2}{2}+\frac{1}{2\sigma^2}$ and  $\kappa\triangleq\frac{\mu^2}{2}+\frac{1}{4\sigma^2}-\frac{\mu^4}{4\alpha^2}$. After extensive analysis given in Appendix \ref{AP2} and profiting from Hypergeometric function, i.e., $_2F_1(a,b;c;d)$ in \cite{HYGEO}, a closed-form expression of (\ref{pEE}) is given as\footnote{See Appendix \ref{AP2} for details.}

\begin{align}
    &P_{e|\textbf{basis}}= 
    \frac{1}{2}-\frac{\mu}{\sqrt{8}}\times
    \frac{_2F_1\left(0,\frac{1}{2};\frac{3}{2};\frac{\mu^2}{\mu^2+\frac{1}{2\sigma^2}}\right)}{\sqrt{\frac{\mu^2}{2}+\frac{1}{4\sigma^2}}} +\frac{\mu}{8\sqrt{2}\alpha\kappa^2\sigma^2}+\nonumber\\
    &\frac{\mu^3}{16\sqrt{2}\alpha^2\kappa^2\sigma^2}\times
    \frac{_2F_1\left(0,\frac{1}{2};\frac{3}{2};\frac{\mu^4}{\mu^2+4\alpha^2\kappa^2}\right)}{\sqrt{\frac{\mu^4}{4\alpha^2}+\kappa^2}}+\nonumber\\
    & \frac{\mu\left(\alpha-\frac{\mu^2}{2\alpha}\right)}{4\sqrt{2}\alpha\kappa^2\sigma^2}\times \frac{_2F_1\left(0,\frac{1}{2};\frac{3}{2};\frac{(\alpha-\frac{\mu^2}{2\alpha})^2}{(\alpha-\frac{\mu^2}{2\alpha})^2+\kappa^2}\right)}{\sqrt{\left(\alpha-\frac{\mu^2}{2\alpha}\right)^2+\kappa^2}}.
     \label{cf}
\end{align}

%% file: Sections/4_modifs.tex
In this section, we present two schemes by modifying the basis scheme in Section \ref{sec2}, which improve the BER and mitigate the interference imposed on original ambient OFDM signal compared to the basis scheme at the cost of data rate, respectively.

\subsection{\textbf{Modification-\rom{1}} }

\input{Subs/2_sub.tex}

 \subsection{\textbf{Modification-\rom{2}} }
\input{Subs/3_sub.tex}

%% file: Subs/2_sub.tex
\label{two}
Since low power IoT devices are mostly accompanied by low/medium data transmission rate (i.e., $<1$ Mbps \cite{AccessSurvey,NB}), instead of transmitting one bit over each subcarrier in the basis scheme, one bit can be sent over a block of subcarriers. It improves the BER performance of the system specifically in frequency-selective channels. 
Furthermore, when $N$ is large enough and the number of paths is limited, i.e., $N\gg \mathfrak{L}$, the channel coefficients of the adjacent subcarriers are being largely correlated \cite{jin-1}. Therefore, in order to further improve the BER performance, an interleaved block transmission can be employed, which increases the diversity gain.
 From \eqref{assu1}, consider $G$ blocks of $L$ subcarriers among a total of $N$ subcarriers of the OFDM spectrum. Then, with block filtering matrix, denoted by $\mathbf{R}_b$, (\ref{eq4}) can be re-expressed as
 \begin{equation}
     \mathbf{y}_b= h_c\,\beta\, \mathbf{R}_b \, \mathbf{u}.
\label{eqb1}
 \end{equation}
 
 Then, $G$ bits are sent over $G$ blocks of $L$ subcarriers using repetition coding (i.e., $L$ is an odd number).
 $G=\lfloor \frac{N}{L}\rfloor$ is the total number of blocks/bits over the frequency spectrum, i.e., $N_d=G$. At the reader, the received signal over the $g^{\text{th}}$ block of subcarriers, similar to (\ref{EQ:r2}), can be expressed as
 \be
\mathbf{r}_g = (\mathbf{\tilde{H}}_{a,g} + \overbrace{h_c\beta \mathbf{\tilde{H}}_{b,g}}^{\mathbf{\tilde{H}}_{s,g}} \mathbf{d}_g) \mathbf{s}_g + \mathbf{\tilde{n}}_g,\quad \ g = 0,\ldots, G-1.
	\label{EQ:r3}    
\ee
Majority logic decoding \cite{jinB2}, along with the simple energy detector given in Subsection \ref{2.A.2}, is used for the signal detection at the reader. Consequently, from (\ref{cf}), the error probability of the \textbf{Modification-\rom{1}} can be obtained by
\begin{equation}
  P_{e|\textbf{{Mod-\rom{1}}}}=\sum_{\ell=\frac{L+1}{2}}^L \left( \begin{array}{c}
          L \\
         \ell 
    \end{array}\right) P_{e|\textbf{basis}}^\ell (1-P_{e|\textbf{basis}})^{L-\ell}.
\end{equation}
\subsection{Performance Analysis of the \textbf{Modification-\rom{1}}}
In the \textbf{Modification-\rom{1}}, the data transmission rate decreases by a factor of $\rho=\frac{G}{N}$, while a better BER performance is achieved. In other words, utilization of repetition coding in a block with $L$ subcarriers can correct up to $\frac{L-1}{2}$ errors in each block.
For examples, considering an OFDM system based on IEEE 802.11a specifications, authors in \cite{of-3} state that $1$ Mbps is a satisfactory data rate for such low data rate applications; accordingly, our proposed method can achieve the same data rate when $G=4$ and $L=13$, but with 13 times higher diversity gain. Also, repetition coding and majority logic decoding are simple processes with a low implementation complexity among linear block codes \cite{RepCode}. Moreover, the complexity of $\mathbf{R}_b$ is the same as $\mathbf{R}$ in (\ref{eq4}). 
Additionally, it can still support a higher data transmission rate by using smaller blocks compared to the existing methods in \cite{of-3,of-0,of,Sensor,of1,F-Guanding,rev3}. In other words, since each block of subcarriers can transmit one bit data, smaller blocks lead to a higher number of blocks, i.e., $G$, when the total number of subcarriers is fixed. It leads to a higher data rate compared to the larger blocks.

%% file: Subs/3_sub.tex
\label{three}
\begin{figure}[t]
\centering
\subfloat[\label{aa}]{
 \includegraphics[width=0.45\linewidth, height=6cm]{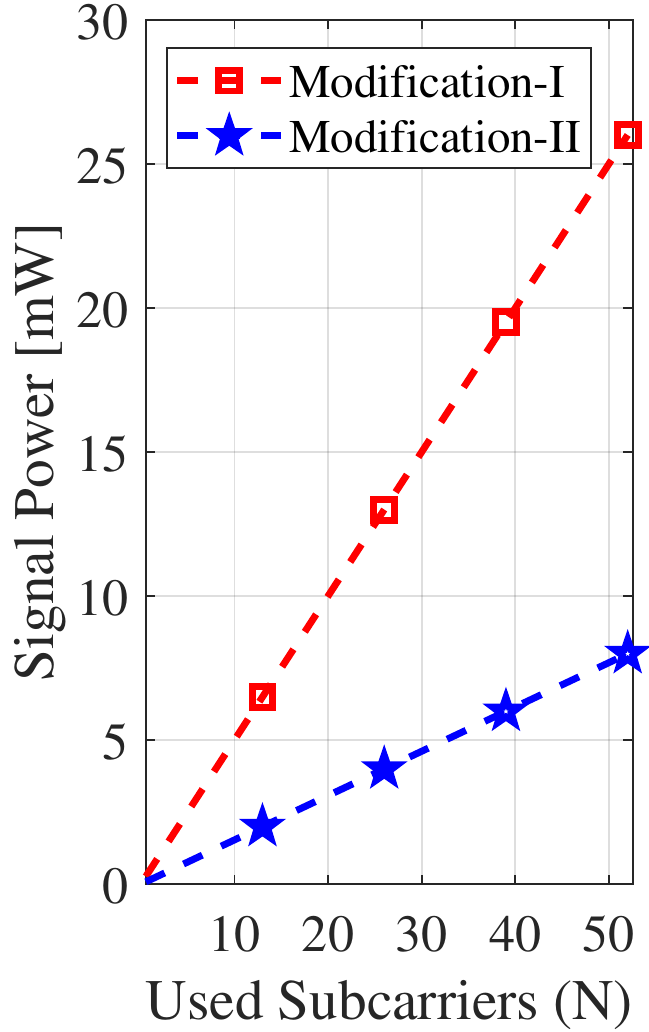}%
 }
 \subfloat[]{
  \includegraphics[width=0.45\linewidth, height=6cm]{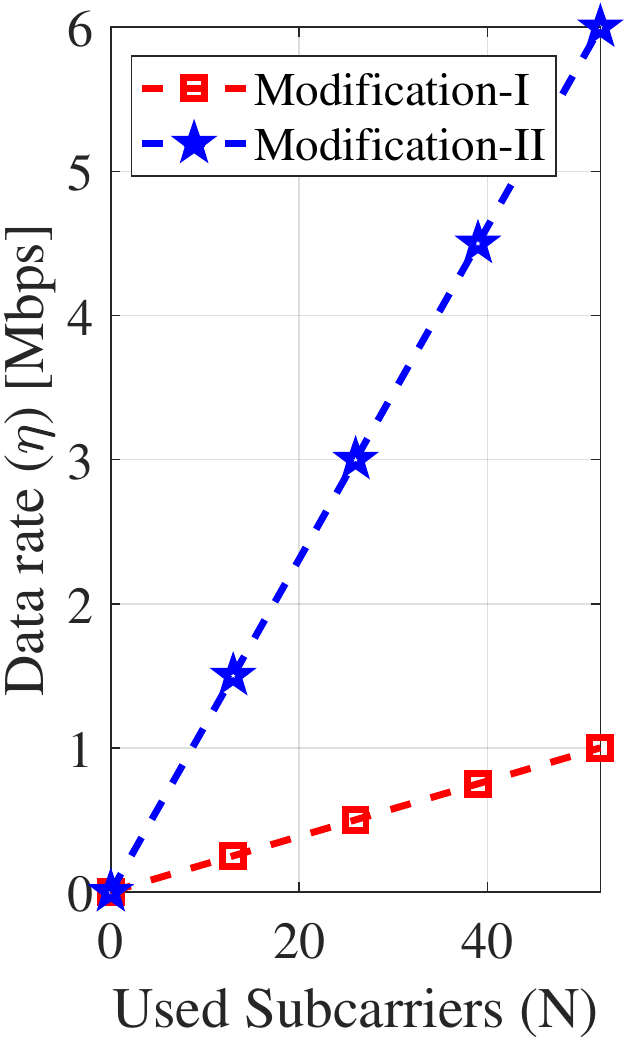}%
  }
 \captionsetup{width=1\linewidth}
 \caption{ Evaluation of signal power and data rate with respect to the number of used subcarriers in \textbf{Modification-\rom{1}} and \textbf{Modification-\rom{2}} when $G=4$, $L=13$, $M=2$, $E_b=1$ mW and $T=4\,\mu$sec.} 
 \label{fig:pow}
\end{figure}
Both of the basis scheme and \textbf{Modification-\rom{1}} have random number of active and inactive subcarriers in the backscattered signal depending on $\mathbf{d}$. Therefore, the power of the signal changes randomly. Assuming $d_l\in\{0,\,1\}$ is equally likely, i.e., $\Pr(d_l=1)=\Pr(d_l=0)=\frac{1}{2}$, there are an average of $\frac{N}{2}$ active subcarriers being backscattered which become interfering subcarriers for a legacy receiver. Note that both the OFDM signal transmitted by the source and the backscattered signal by the tag occur at the same frequency. Therefore, to control the power of the backscattered signal and keep the interference level for the legacy receiver low, it would be desirable to use a fraction of $N$ subcarriers, say $M(\ll N)$ active subcarriers and $K(=N-M)$ null subcarriers for the backscatter communication. 
To this end, we can consider IM technique \cite{IM} in \textbf{Modification-\rom{2}}. As a result, a fixed number of $Q=G\,M$ active subcarriers are being backscattered. Fig. \ref{fig:pow} (a) shows that IM backscattered signal in \textbf{Modification-\rom{2}} has much lower power than the backscattered signal in the \textbf{Modification-\rom{1}} when $M=2$ leading to lower interference level for legacy receiver. On the other hand,  Fig. \ref{fig:pow} (b) shows that better data rate performance is achieved as well.

In particular, we assume an IM model where only the indices of active subcarriers are used to send information. Let the $g^{\text{th}}$ IM block, denoted by $\mathbf{b}_g=[b_{g,0}, \cdots , b_{g,L-1}]^T$, contain $M$ active subcarriers and $K$ inactive subcarriers, i.e., $||\mathbf{b}_g||_0=M$. For convenience, let $I_g$ denote the support of $\mathbf{b}_g$, i.e., $I_g=\{l|[\mathbf{b}_g]_l=1\}$. Therefore, the total number of bits transmitted per OFDM symbol duration is given by
\begin{equation}
N_b=G\times\lfloor \log_2 \binom{L}{M} 
\rfloor.
\label{eqNb}
\end{equation}
Note that $N_b$ in IM block transmission is equal to or greater than $G$ in \textbf{Modification-\rom{1}} and is less than $N$ in basis scheme. 
From (\ref{EQ:r3}), the $g^{\text{th}}$ IM block can be rewritten as
 \be
\mathbf{r}_g = (\mathbf{\tilde{H}}_{a,g} + \overbrace{h_c\beta \mathbf{\tilde{H}}_{b,g}}^{\mathbf{\tilde{H}}_{s,g}} \mathbf{b}_g) \mathbf{s}_g + \mathbf{\tilde{n}}_g,\quad \ g = 0,\ldots, G-1.
	\label{EQ:r4}
\ee
For the ML detection, we have
\begin{equation}
    \mathbf{\hat{b}}_g=\underset{\mathbf{b}_g}{\mathrm{argmin}}\biggm| ||\mathbf{r}_g||^2-||\mathbf{\tilde{H}}_{a,g}+\mathbf{\tilde{H}}_{s,g}\mathbf{b}_g||^2E_b\biggm|.
\end{equation}
Since the analysis of ML detection of such OFDM-IM blocks is well studied in \cite{basar,jin-1,IM2,choi2017,IM}, we do not include details here.

\subsection{Performance Analysis of the \textbf{Modification-\rom{2}}}
Since backscattered signal uses the same spectrum as the legacy system, it can cause an interference over the legacy system. The average backscattered signal power in the basis scheme and \textbf{Modification-\rom{1}} is given by
\begin{equation}
    \text{Backscattered Signal Power}=\frac{\beta^2 E_b\sum_{l\in I_1}|\tilde{h}_{b,l}|^2}{N},
    \label{sigpow}
\end{equation}
where $I_1$ denotes the support for indices of active subcarriers. Considering $\Pr(d_l=1)=\Pr(d_l=0)=\frac{1}{2}$, we have  an average of $\mathbb{E}\left\{||I_1||_0\right\}=\frac{N}{2}$ active subcarriers. On the other hand, in \textbf{Modification-\rom{2}}, the total number of active subcarriers is fixed and $||I_1||_0=Q$, i.e., $Q=GM$. Therefore, in general, from \eqref{sigpow}, the signal power in \textbf{Modification-\rom{2}} is fixed while two former schemes might have variable signal power in consecutive backscattered symbols with respect to the forward-link channel coefficients in $\tilde{\mathbf{h}}_{b}$. Moreover, \textbf{Modification-\rom{2}} leads to a lower interference level for the legacy system when $Q<\frac{N}{2}$ leading to a lower number of active subcarriers compared to the two former schemes. Furthermore, from \eqref{eqNb}, \textbf{Modification-\rom{2}} supports higher data rate compared to the \textbf{Modification-\rom{1}} when $M>1$.

%% file: Sections/5_simul.tex
In this section, we provide simulation results to evaluate the performance of the proposed schemes.
We consider an OFDM system model based on IEEE 802.11a specifications. The parameters used in simulations are given in Table \ref{tab:par}, unless otherwise
specified.
\begin{figure}[t]
\centering
\captionsetup{width=1\linewidth}
  \includegraphics[width=1\linewidth, height=7cm]{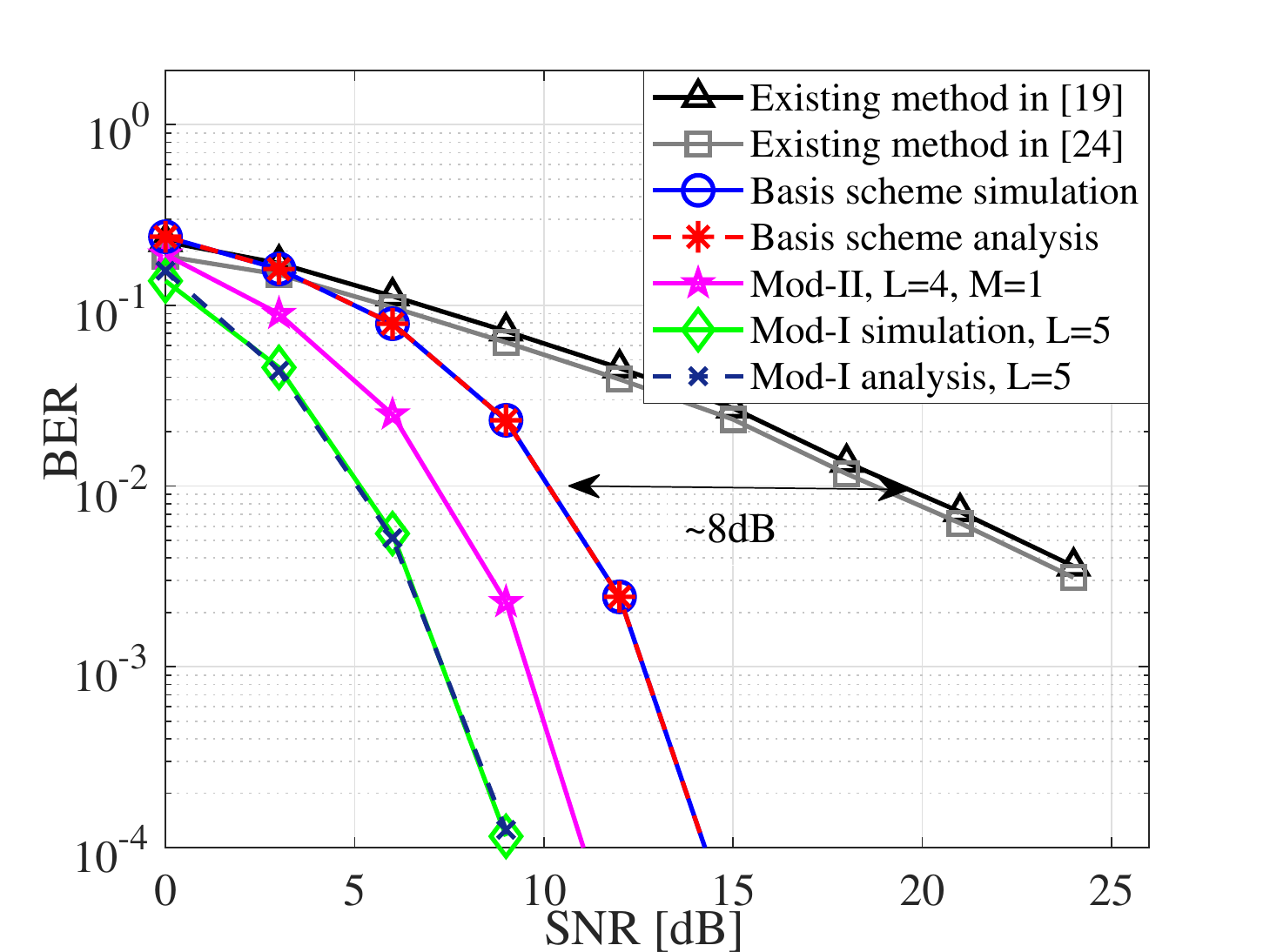}%
  \caption{BER performance in a short-range communication with strong LoS link when $\tau$=0.}%
  \label{fig:22}
\end{figure}
\begin{table}[h]
\begin{center}
\centering
\captionsetup{width=1\linewidth}
\caption{System numerical parameters.}
\label{tab:par}
\begin{tabular}{cc} \hline
\hline
System parameters & Corresponding value\\ 
 \hline
 DFT size & 64\\
 Number of used subcarriers & 52\\
 DFT sampling frequency & $20$ MHz\\
 Subcarrier spacing & $312.5$ kHz\\
 Cyclic prefix, $T_{cp}$ & $0.8\,\mu$sec\\
 Total OFDM symbol duration, $T_s$ & $4\,\mu$sec\\
 \hline
\hline
\end{tabular}
\end{center}
\end{table}

\subsection{BER Performance Comparison}
For the first evaluation, we compare the BER performance of our proposed schemes with that of the existing methods in \cite{of-0} and \cite{F-Guanding} which are the baseline methods for other existing methods in \cite{of,Sensor,of1}. The BER performance of the existing method in \cite[Eq.(54)]{of-0} is a nonlinear function of $\gamma$. The input of the function in a high SNR is as follows:
\begin{equation}
    \psi\approx\sqrt{\frac{\left(N_{cp}-\mathfrak{L}\right)\left(\gamma^2-\gamma\sqrt{1+\frac{2\ln(\gamma)}{N_{cp}-\mathfrak{L}}}\right)^2+2\ln(\gamma)}{\gamma^4}}.
    \label{psi}
\end{equation}

As SNR increases, $\psi$ approaches $\sqrt{ (N_{cp}-\mathfrak{L})+2\ln(\gamma)}$. Therefore, the BER decreases slowly. It is evident that its BER probability depends on
the portion of CP not affected by the multipath channel, i.e., $N_{cp}-\mathfrak{L}$. Thus, the performance of the existing method in \cite{of-0} becomes worse if the maximum channel delay spread, i.e., $\tau$, increases. Moreover, it requests an extra overhead since it needs longer CP than the maximum channel delay spread.
In order to show the robustness of our proposed method to the maximum channel delay spread compared to the existing method in \cite{of-0}, we consider two different values of $\tau=0$ and $\tau=T_{cp}$ in our simulations.
 On the other hand, the BER performance of the existing method in \cite{F-Guanding} depends on the number of null subcarriers in an OFDM symbol. Although the robustness of the existing model in \cite{F-Guanding} to the maximum channel delay spread is higher than that in \cite{of-0}, its performance degrades when the number of null subcarriers is low.

\begin{figure}[t]
\centering
\captionsetup{width=1\linewidth}
\includegraphics[width=1\linewidth, height=7cm]{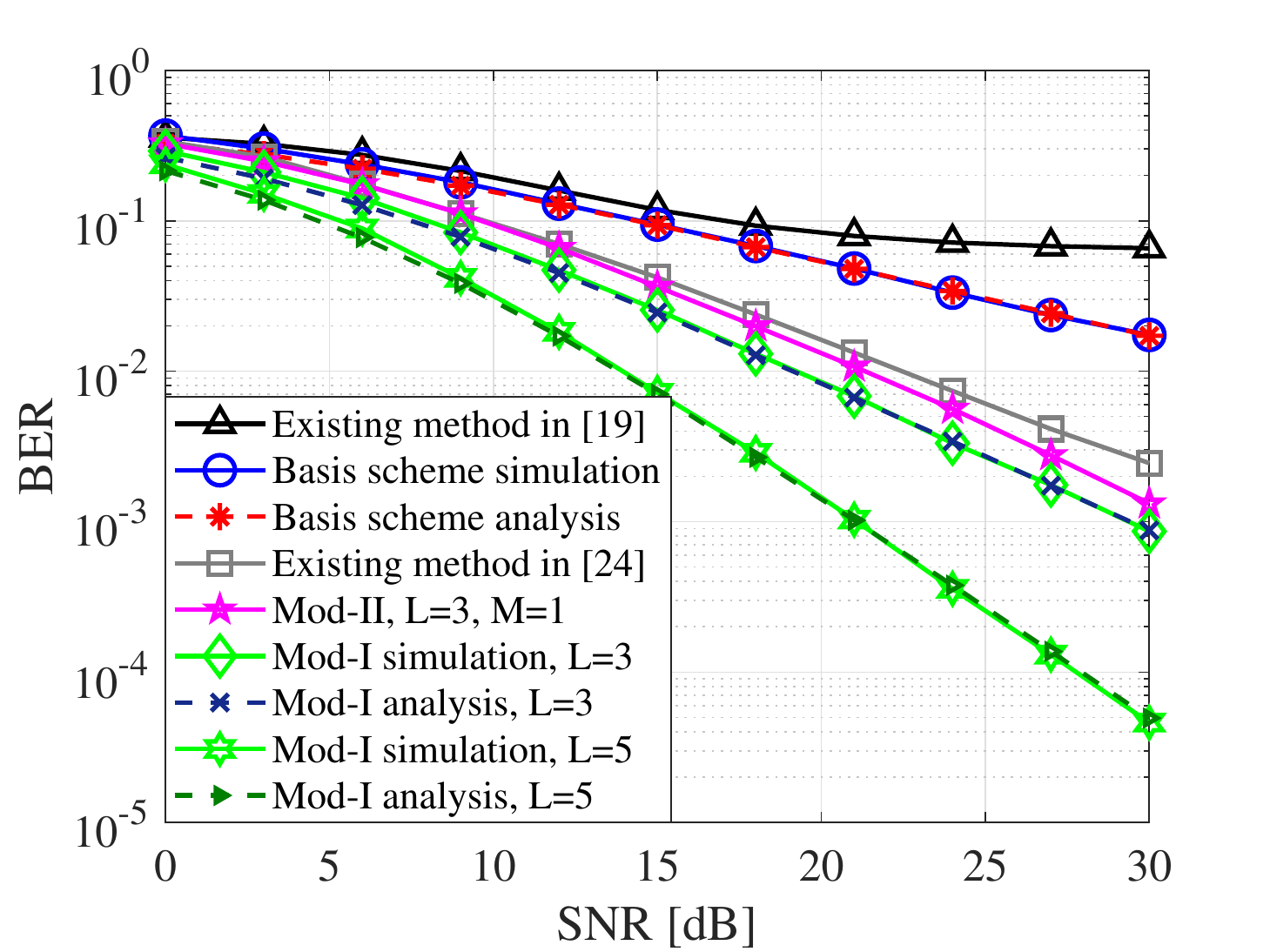}%
  \caption{BER performance in a frequency-selective channel ($\tau=T_{cp}$).}%
  \label{fig:2}
\end{figure}
Fig. \ref{fig:22} compares the BER performance of the proposed AmBC schemes and the existing methods in \cite{of-0} and \cite{F-Guanding} in a short-range and poor-scattering environment such as those considered in \cite{chan2,5,wcnc}. In fact, the
maximum delay spread is less than the sampling period due to
a short-distance scenario, i.e., $\tau=0$ \cite{5}. Thus, the paths are considered to be one resolvable path with strong LoS, e.g., AWGN channel. 
It is shown that the proposed schemes outperform the existing methods in \cite{of-0} and \cite{F-Guanding} (e.g., about 8 dB SNR gain at BER$=10^{-2}$ for the basis scheme). Moreover, \textbf{Modification-\rom{2}} and \textbf{Modification-\rom{1}} have respectively better BER performances than the basis scheme.

On the other hand, Fig. \ref{fig:2} depicts their BER evaluations over a frequency-selective channel. A Rayleigh fading multipath channel with $\tau=T_{cp}$ is considered for this simulation. As shown in Fig. \ref{fig:2}, the existing method in \cite{of-0} has a poor performance when $N_{cp}-\mathfrak{L}$ becomes zero. However, our proposed schemes significantly improve the BER performance of the system. The existing model in \cite{F-Guanding} performs better than the existing model in \cite{of-0} and the basis scheme. But it is also shown that by increasing the size of the blocks in \textbf{Modification-\rom{1}}, a better BER performance can be achieved in our proposed method. \\

\subsection{Data Rate and Power Evaluation}
\begin{figure}[t]
\centering
\captionsetup{width=1\linewidth}
  \includegraphics[width=9cm, height=7cm]{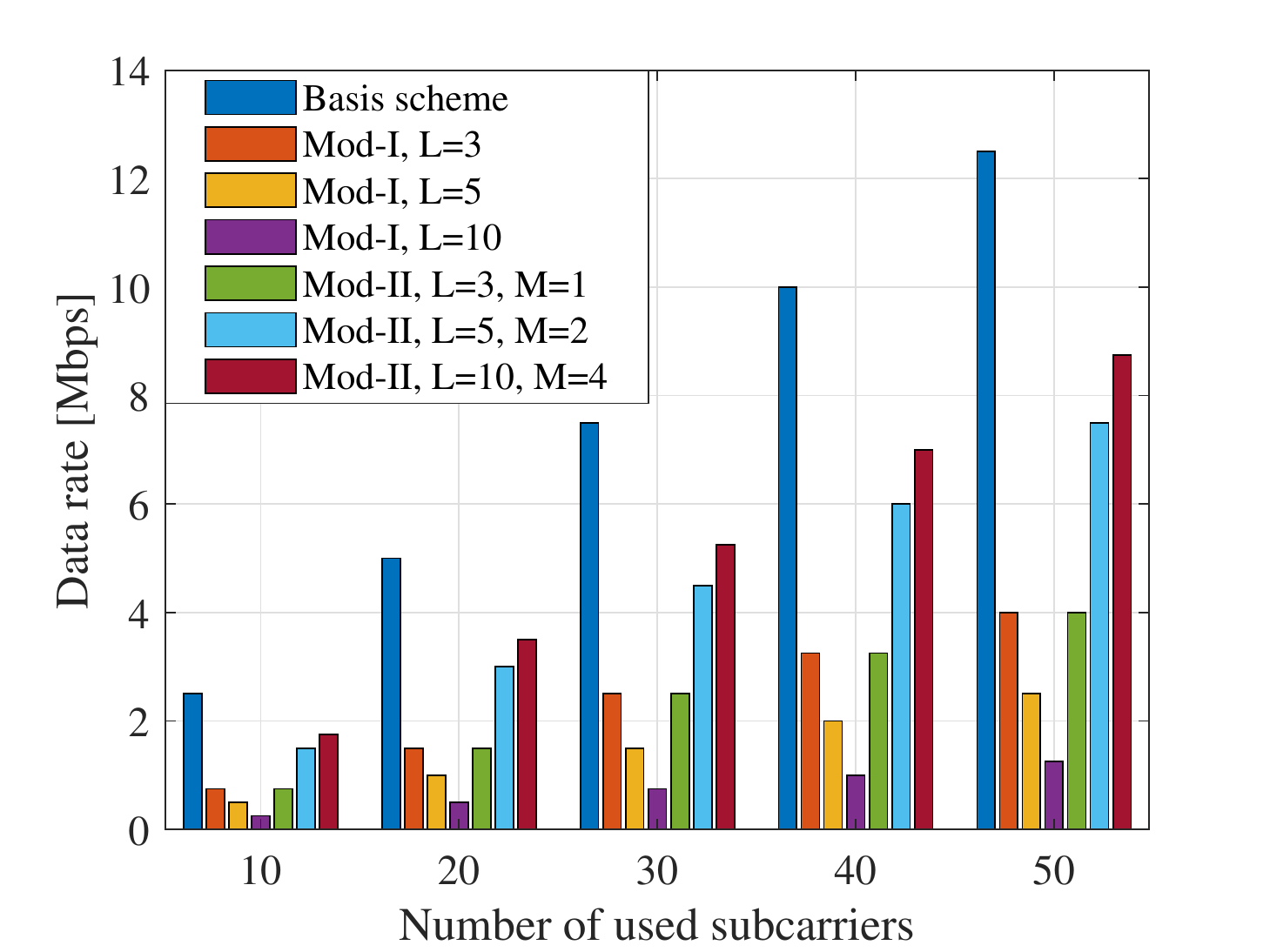}%
  \caption{Data rate evaluation of the proposed schemes.}%
  \label{fig:9}
\end{figure}

\begin{figure}[t]
\centering
\captionsetup{width=1\linewidth}
  \includegraphics[width=9cm, height=7cm]{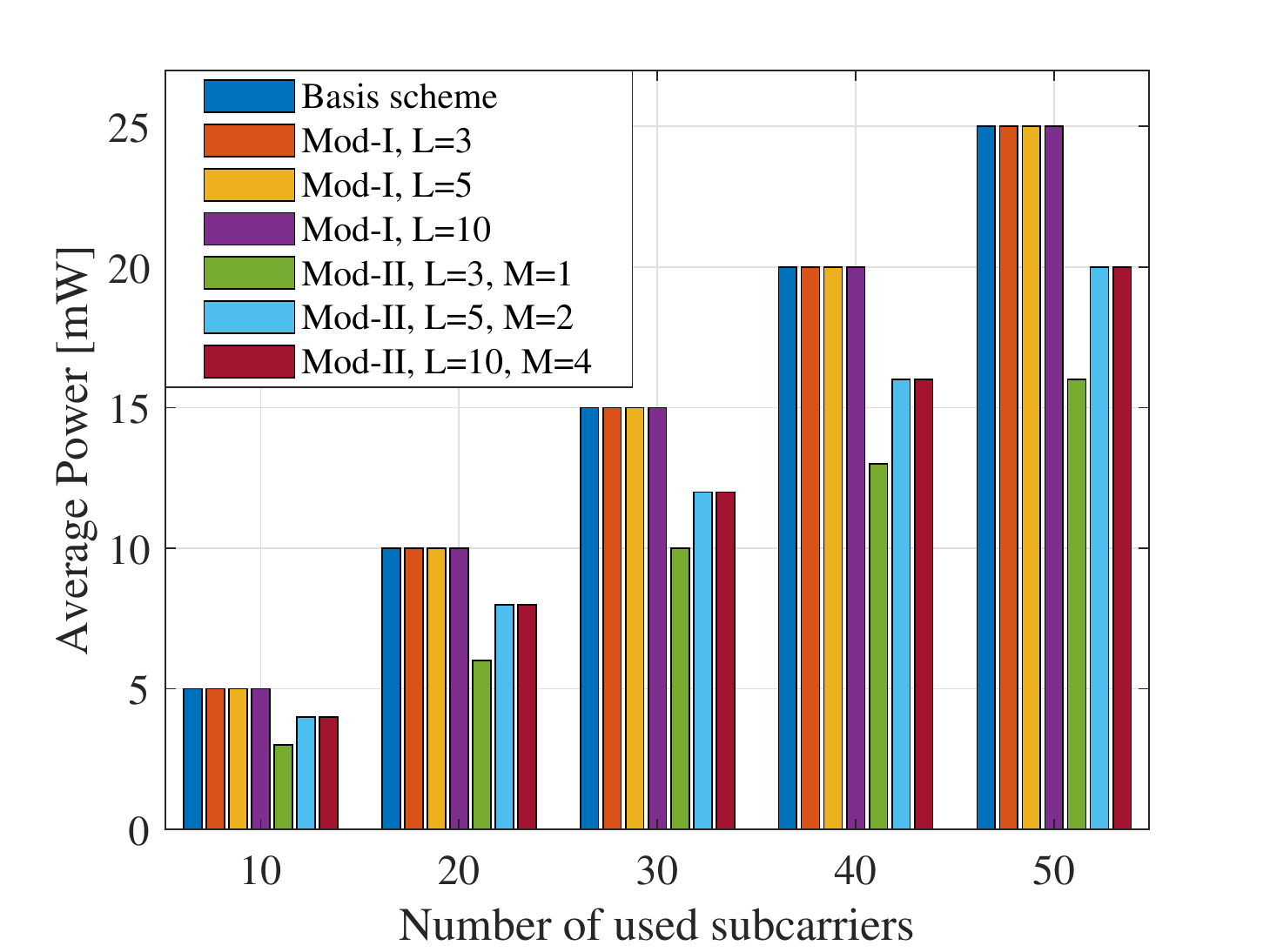}%
  \caption{Signal power evaluation of the proposed schemes, $E_b=1$ mW.}%
  \label{fig:10}
\end{figure}

In this subsection, we evaluate the performance of the proposed method in terms of data rate while observing their transmitted signal power. Fig. \ref{fig:9} shows the data rate of the proposed basis scheme and its modifications with respect to the number of used subcarriers. It is evident that the data rate increases when the number of used subcarriers increases.
In the basis scheme, the data rate is $\eta = \frac{N}{T_s}$. In \textbf{Modification-\rom{1}}, the data rate is $\eta=\frac{G}{T_s}=\lfloor\frac{N}{L}\rfloor\frac{1}{T_s}$ and decreases as $L$ increases. Finally, the data rate in \textbf{Modification-\rom{2}}, from (\ref{eqNb}), is $\eta=\frac{N_b}{T_s}$.
Furthermore, as shown in Fig. \ref{fig:10}, the average signal power for the basis scheme and \textbf{Modification-\rom{1}} are the same. Additionally, \textbf{Modification-\rom{2}} supports higher data rate compared to \textbf{Modification-\rom{1}} while enjoying lower signal power.

\begin{figure*}[t]
\centering
\captionsetup{width=0.3\linewidth}
\subfloat[The BRI ratio in the basis scheme.]{%
 \includegraphics[width=0.31\linewidth, height=5.5cm]{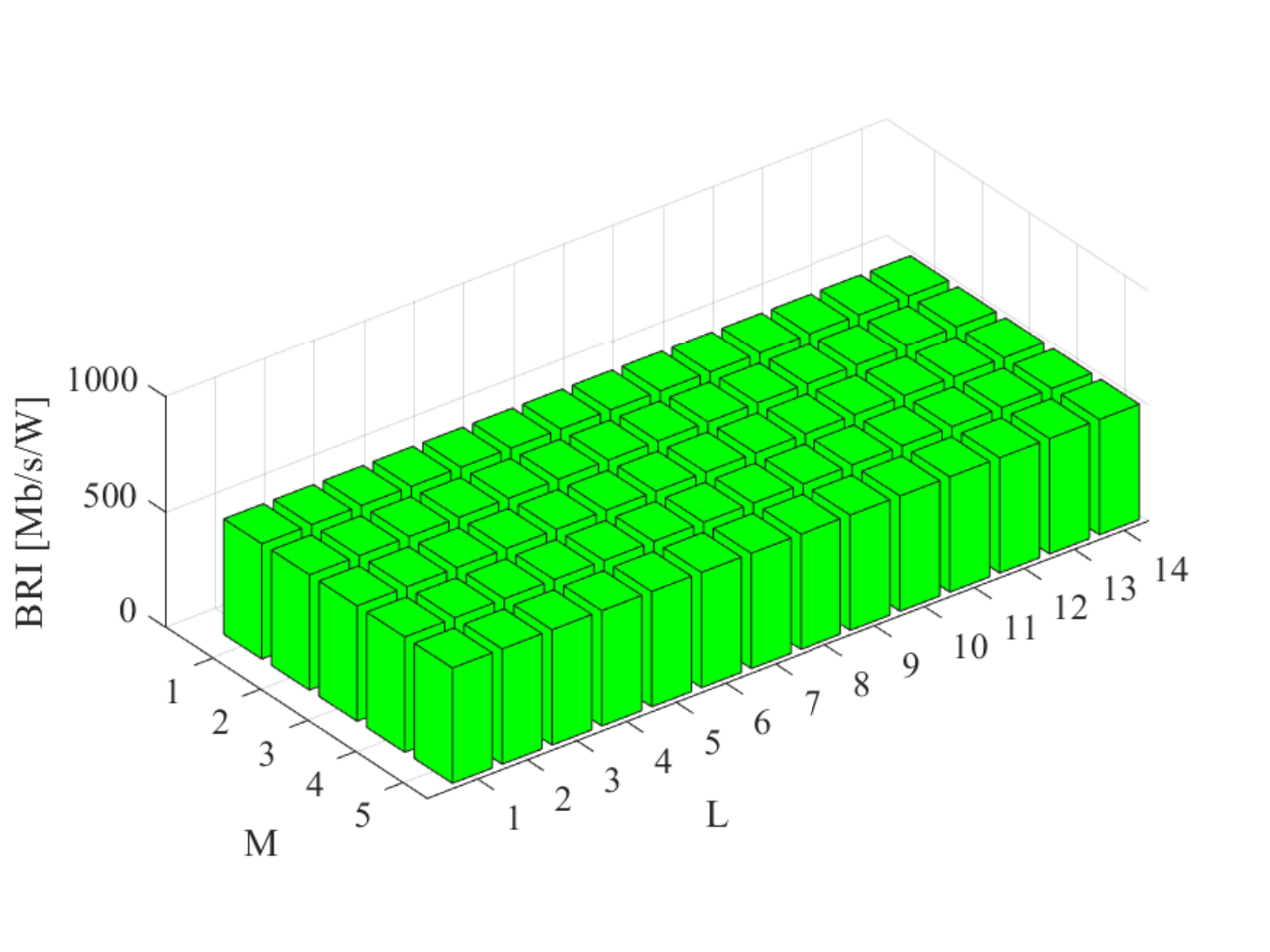}%
 } 
 \hspace{0.1cm}
\captionsetup{width=0.3\linewidth}
\subfloat[The BRI ratio in \textbf{Modification-\rom{1}}.]{%
 \includegraphics[width=0.31\linewidth, height=5.5cm]{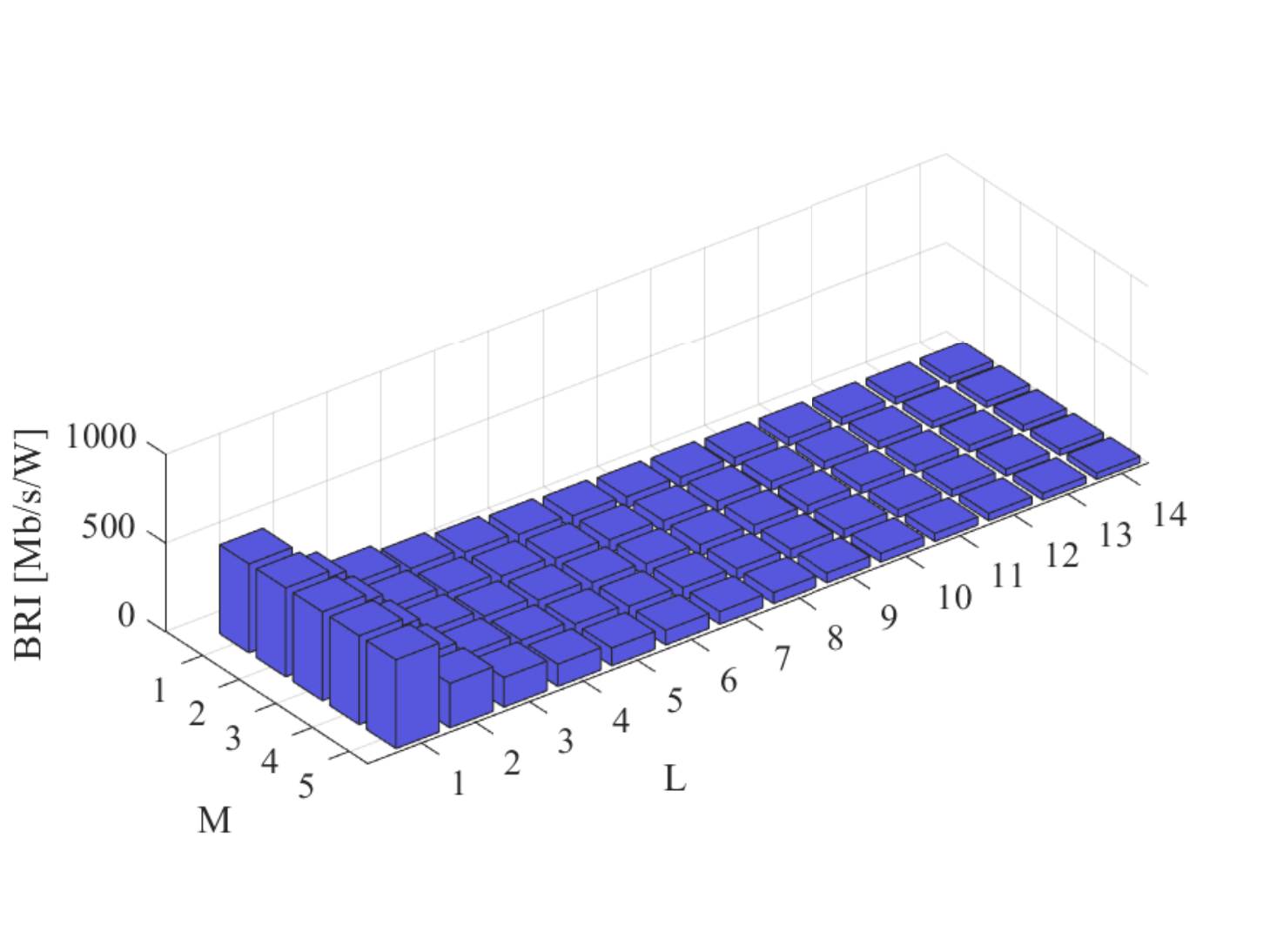}%
 } 
  \hspace{0.1cm}
\captionsetup{width=0.3\linewidth}
\subfloat[The BRI ratio in \textbf{Modification-\rom{2}}.]{%
 \includegraphics[width=0.31\linewidth, height=5.5cm]{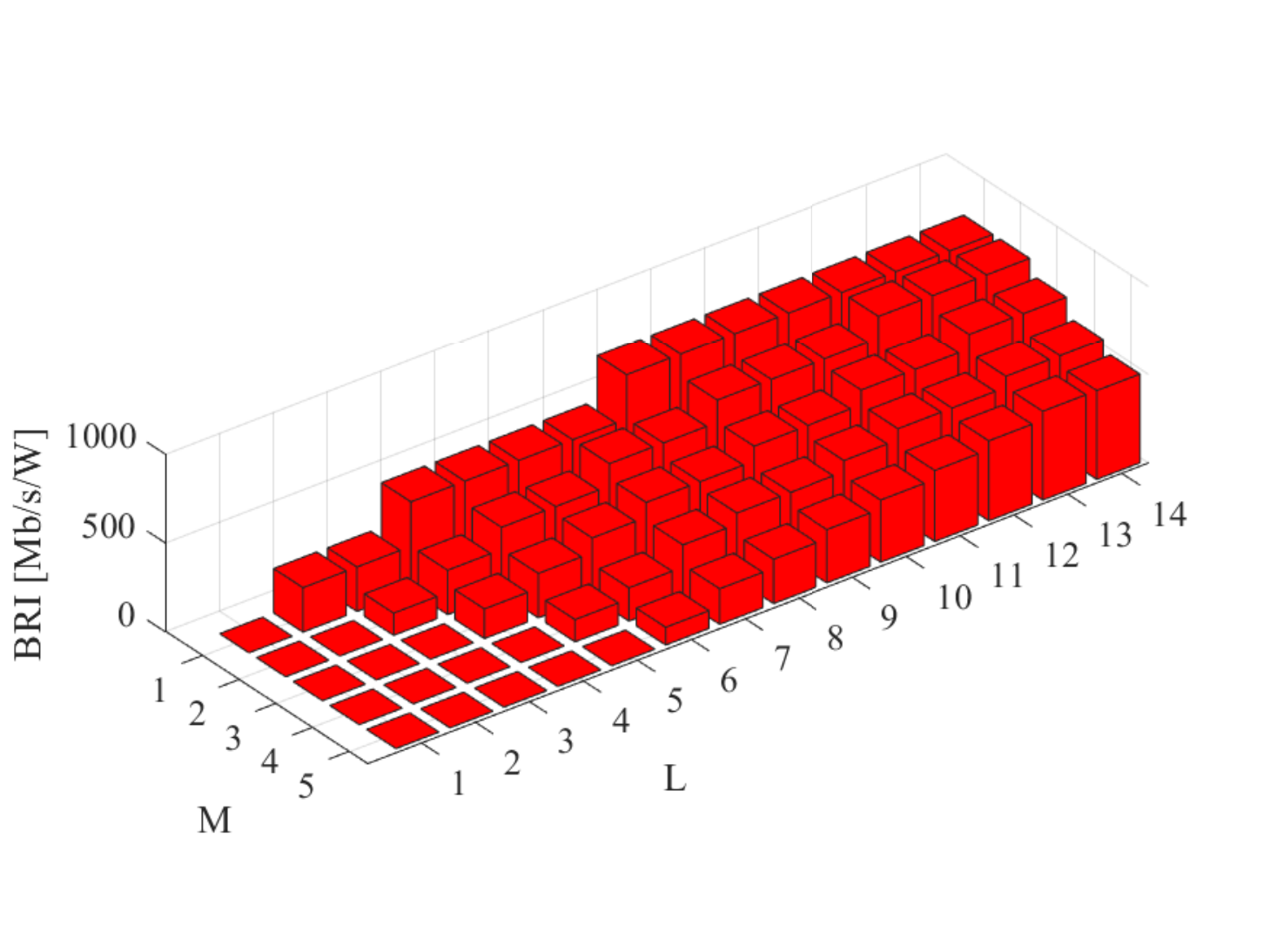}%
 } 
 \captionsetup{width=1\linewidth}
 \caption{ BRI ratio evaluation of the proposed basis scheme and its modifications. The BRI in the basis scheme is independent of $M$ and $L$. The BRI in \textbf{Modification-\rom{1}} depends on $L$. The BRI in \textbf{Modification-\rom{2}} depends on both of $M$ and $L$.}
 \label{fig:11}
\end{figure*}

The power of the backscattered signal can cause an interference over the ambient OFDM signal for the legacy receiver. Hence, we define the bit-rate-to-interference (BRI) ratio as the data rate, i.e., $\eta$, over the interference produced by active subcarriers. Let $\lambda$ denote the BRI ratio and $E_b$ is set to 1 mW. In the basis scheme, considering $\eta=\frac{N}{T_s}$ transmitted bits over an average of $\frac{N}{2}$ active subcarriers\footnote{$d_l$ is equally likely, i.e., $\Pr(d_l=0)=\Pr(d_l=1)=\frac{1}{2}$.}, the BRI is given by
\begin{equation}
    \lambda_{\textbf{basis}}=\left.\frac{\frac{N}{T_s}}{\frac{N}{2}\, E_b}\right.=500  \,Mb/s/W.
\end{equation}
The green columns in Fig. \ref{fig:11} (a) show the fixed BRI for the basis scheme. For \textbf{Modification-\rom{1}} where each bit is transmitted by a block of subcarriers, we have
\begin{equation}
    \lambda_{\textbf{Mod-\rom{1}}}=\left.\frac{\frac{G}{T_s}}{\frac{G\, L}{2}\, E_b}\right.=\frac{500}{L} Mb/s/W.
\end{equation}
Fig. \ref{fig:11} (b) shows that by increasing the size of blocks, i.e., $L$, the BRI decreases. Moreover, for \textbf{Modification-\rom{2}} where IM is employed for backscatter communication, from (\ref{eqNb}), we have
\begin{align}
    \lambda_{\textbf{Mod-\rom{2}}}&=\left.\frac{\frac{G}{T_s}\lfloor \log_2\left(\begin{array}{c}
         L  \\
         M 
    \end{array}\right)\rfloor}{G\, M\, E_b}\right.\nonumber\\
    &=\frac{\lfloor \log_2\left(\begin{array}{c}
         L  \\
         M 
    \end{array}\right)\rfloor \times 250}{M} Mb/s/W.
\end{align}
 Fig. \ref{fig:11} (c) shows that the highest BRI of the \textbf{Modification-\rom{2}} is when $L$ increases and $M$ decreases.
 
 

%% file: Sections/Conclusions.tex
In this paper, we studied subcarrier-wise backscatter communication over ambient OFDM in the frequency domain and proposed a basis scheme and its two modifications. In the basis scheme, OOK modulation for each subcarrier has been used to transmit one bit per subcarrier. Considering interleaved block transmission, it was shown that \textbf{Modification-\rom{1}} improves the BER performance of the system significantly. 
Finally, in \textbf{Modification-\rom{2}}, IM was employed to control the power of the signal so that the interference imposed on the original OFDM signal can be reduced.
Moreover, a closed-form optimal decision threshold has been derived for the amplitude of the received subcarrier to detect the data at the reader. Analytical expressions for the BER probability was also derived for both AWGN and Rayleigh fading channels which coincide with the simulation results. 


%% file: Subs/Ap1.tex
In order to find $\Pr(|r|>\delta\, \big|d=0,A<B,B)$, we have to solve the following integral with respect to Fig. \ref{fig:LIFreq} (a)
\begingroup\makeatletter\def\f@size{9}\check@mathfonts
\begin{align}
   \Pr(|r|>&\delta\,  \big|d=0,A<B,B)=\nonumber\\
   &\frac{1}{2\sqrt{\pi N_0}}\biggl[ \int_{-\infty}^{-\delta} e^{-\frac{(x+A\sqrt{E_b})^2}{N_0}} dx+\int^{\infty}_{\delta} e^{-\frac{(x+A\sqrt{E_b})^2}{N_0}} dx\nonumber\\ 
    &+\int_{-\infty}^{-\delta} e^{-\frac{(x-A\sqrt{E_b})^2}{N_0}} dx+\int^{\infty}_{\delta} e^{-\frac{(x-A\sqrt{E_b})^2}{N_0}} dx\biggr].
\end{align}
\endgroup
Therefore, we have
\begin{align}
   \Pr(|r|>\delta\,  &\big|d=0,A<B,B)=\nonumber\\
   &\mathcal{Q}\left( (B-A)\sqrt{\frac{\gamma}{2}}\right)+\mathcal{Q}\left((3A+B) \sqrt{\frac{\gamma}{2}}\right)
   \label{aeq1}.
\end{align}
Likewise we have,
\begin{align}
   \Pr(|r|<\delta\,  &\big|d=1,A<B,B)=\nonumber\\
   &\mathcal{Q}\left( (B-A)\sqrt{\frac{\gamma}{2}}\right)-\mathcal{Q}\left((3B+A) \sqrt{\frac{\gamma}{2}}\right)
   \label{aeq2}.
\end{align}

Since, $\mathcal{Q}(x) \leqslant \exp(-\frac{x^2}{2})$ and $A,B  \in \mathbb{R}^+$ \footnote{$R^+=\{x|x\geqslant 0, x\in \mathbb{R}\}$}, as SNR increases, we can see $ \mathcal{Q}\left( (B-A)\sqrt{\frac{\gamma}{2}}\right)\gg \mathcal{Q}\left((3A+B) \sqrt{\frac{\gamma}{2}}\right)$ and $\mathcal{Q}\left( (B-A)\sqrt{\frac{\gamma}{2}}\right) \gg \mathcal{Q}\left((3B+A) \sqrt{\frac{\gamma}{2}}\right)$ in (\ref{aeq1}) and (\ref{aeq2}), respectively. 
In other words, the effect of $\mathcal{Q}\left((3A+B) \sqrt{\frac{\gamma}{2}}\right)$ and $\mathcal{Q}\left((3B+A) \sqrt{\frac{\gamma}{2}}\right)$ on BER probability in (\ref{aeq1}) and (\ref{aeq2}) is negligible and we ignore them in (\ref{eqAP1}) in Section \ref{SBER}. The similar analysis is considered for (\ref{eqap12}) in Section \ref{SBER}.

%% file: Subs/AP2.tex
In order to find the closed-form expression of (\ref{pEE}), we first set $\sigma_a^2=\sigma_b^2=\sigma^2$, $\sigma_c^2=1$, $\mu=\sqrt{\frac{\gamma}{2}}$, and $\beta=1$ to simplify the notation and analysis. Thus, we have
\begin{align}
    \left\{\begin{array}{l}
     h_a\sim \mathcal{CN}(0, \sigma^2)\\
     \vspace{1mm}
     h_a+h_s \sim \mathcal{CN} (0, 2\sigma^2 )
     \end{array}\right. .
\end{align}
Therefore, we can assume that $A$ and $B$ are Rayleigh random variables with average power of $\sigma^2$ and $2\sigma^2$, respectively \cite{choi2017,GFIM}. Consequently, the error probability of the basis scheme in (\ref{pEE}) is given by
\begingroup\makeatletter\def\f@size{8}\check@mathfonts
\begin{align}
    &P_{e|\textbf{basis}}=\nonumber\\
    &\overbrace{\int_0^\infty  \int_0^B \mathcal{Q}\left [(B-A)\mu \right] \frac{A}{\sigma^2}\exp \left(-\frac{A^2}{2\sigma^2}\right)  \frac{B}{2\sigma^2}\exp \left(-\frac{B^2}{4\sigma^2}\right)\, dA\, dB }^{f_1(\mu)}+\nonumber\\ 
    &
    \overbrace{\int_0^\infty  \int_B^\infty \mathcal{Q}\left [(A-B)\mu \right] \frac{A}{\sigma^2}\exp \left(-\frac{A^2}{2\sigma^2}\right)  \frac{B}{2\sigma^2}\exp \left(-\frac{B^2}{4\sigma^2}\right)\, dA \,dB}^{f_2(\mu)} .
    \label{eq1ap2}
\end{align}  
\endgroup
In order to calculate the first expression of (\ref{eq1ap2}), i.e., $f_1(\mu)$, by taking a partial integration, we have
\begingroup\makeatletter\def\f@size{9}\check@mathfonts
\begin{align}
&f_1(\mu)=\int_0^\infty  \int_0^B \overbrace{\mathcal{Q}\left [(B-A)\mu \right]}^{u}\overbrace{\left[ -\frac{d}{dA}\left(\exp\left(-\frac{A^2}{2\sigma^2}\right)\right)\right]\,dA  }^{v}\nonumber\\
    & \hspace{4.1cm} \times\frac{B}{2\sigma^2}\exp \left(-\frac{B^2}{4\sigma^2}\right)\, dB, 
    \end{align}
    \endgroup
    and consequently
    \begingroup\makeatletter\def\f@size{9}\check@mathfonts
    \begin{align}
    f_1(\mu)=\int_0^\infty  \left[uv\bigg|_0^B-\int_0^Bvdu\right] 
    \frac{B}{2\sigma^2}\exp \left(-\frac{B^2}{4\sigma^2}\right)\, dB.  
    \end{align}
    \endgroup
    We set $\alpha \triangleq \frac{\mu^2}{2}+\frac{1}{2\sigma^2}$, then, we have
    \begingroup\makeatletter\def\f@size{9}\check@mathfonts
    \begin{align}
    f_1(\mu)&=\int_0^\infty \left[\mathcal{Q}(B\mu) - \frac{1}{2} \exp \left(-\frac{B^2}{2\sigma^2}\right)
    +\frac{\mu}{\sqrt{2\pi}}
    \exp\left(-\frac{\mu^2}{2}B^2\right)\right.\nonumber\\
    &\times \left.\underbrace{\int_0^B \exp \left(-\alpha^2 A^2+\mu^2BA\right)\,dA}_{g(B)}
    \right]
    \frac{B}{2\sigma^2}\exp \left(-\frac{B^2}{4\sigma^2}\right)\, dB.  \label{m_s}
    \end{align}
    \endgroup
    Expression of $g(B)$ in (\ref{m_s}) is given by
    \begingroup\makeatletter\def\f@size{9}\check@mathfonts
    \begin{align}
    g(B)=\frac{\sqrt{\pi}}{2\alpha} \exp\left(\frac{\mu^4}{4\alpha^2}B^2\right)\left\{\erf\left(\frac{\mu^2}{2\alpha}B\right)+\erf\left((\alpha-\frac{\mu^2}{2\alpha})B\right)\right\},
    \label{eq2ap2}
    \end{align}
    \endgroup
    where $\erf(x)=\frac{1}{\sqrt{\pi}}\int_{-x}^xe^{-t^2}dt$.
    By substituting (\ref{eq2ap2}) into (\ref{m_s}) we have
        \begingroup\makeatletter\def\f@size{7.5}\check@mathfonts
    \begin{align}
   & f_1(\mu)=\int_0^\infty \mathcal{Q}\left[B\mu\right]\frac{B}{2\sigma^2}\exp \left(-\frac{B^2}{4\sigma^2}\right)\, dB-\nonumber\\
    &\frac{1}{4\sigma^2}\int_0^\infty \exp \left(-\frac{B^2}{2\sigma^2}\right) B \exp \left(-\frac{B^2}{4\sigma^2}\right) \, dB +\nonumber\\
    &\frac{\mu}{4\sqrt{2}\alpha\sigma^2} \int_0^\infty B \exp \left(-B^2\left[\frac{\mu^2}{2}+\frac{1}{4\sigma^2}-\frac{\mu^4}{4\alpha^2}\right]\right) \erf\left(\frac{\mu^2}{2\alpha}B\right) \,dB +\nonumber\\
    &\frac{\mu}{4\sqrt{2}\alpha\sigma^2} \int_0^\infty B \exp \left(-B^2\left[\frac{\mu^2}{2}+\frac{1}{4\sigma^2}-\frac{\mu^4}{4\alpha^2}\right]\right) \erf\left( (\alpha-\frac{\mu^2}{2\alpha})B\right) \,dB.\label{eq24}
\end{align} 
\endgroup

Let define $\kappa\triangleq\frac{\mu^2}{2}+\frac{1}{4\sigma^2}-\frac{\mu^4}{4\alpha^2}$. Then, profiting from Hypergeometric function, i.e., $_2F_1(a,b;c;d)$ in \cite{HYGEO}, (\ref{eq24}) is solved as 
\begin{align}
     f_1(\mu)&=\frac{1}{2}-\frac{\mu}{\sqrt{8}}\frac{_2F_1\left(0,\frac{1}{2};\frac{3}{2};\frac{\mu^2}{\mu^2+\frac{1}{2\sigma^2}}\right)}{\sqrt{\frac{\mu^2}{2}+\frac{1}{4\sigma^2}}} - \frac{1}{6}+\nonumber\\
    &\frac{\mu^3}{16\sqrt{2}\alpha^2\kappa^2\sigma^2}\frac{_2F_1\left(0,\frac{1}{2};\frac{3}{2};\frac{\mu^4}{\mu^2+4\alpha^2\kappa^2}\right)}{\sqrt{\frac{\mu^4}{4\alpha^2}+\kappa^2}}+\nonumber\\
    &\frac{\mu\left(\alpha-\frac{\mu^2}{2\alpha}\right)}{8\sqrt{2}\alpha\kappa^2\sigma^2}\frac{_2F_1\left(0,\frac{1}{2};\frac{3}{2};\frac{(\alpha-\frac{\mu^2}{2\alpha})^2}{(\alpha-\frac{\mu^2}{2\alpha})^2+\kappa^2}\right)}{\sqrt{\left(\alpha-\frac{\mu^2}{2\alpha}\right)^2+\kappa^2}}.
\end{align}

Likewise, $f_2(\mu)$ in (\ref{eq1ap2}) can be obtained as
\begin{align}
     &f_2(\mu)=\frac{1}{6}+ \nonumber\\
     &
     \frac{\mu\left(\alpha-\frac{\mu^2}{2\alpha}\right)}{8\sqrt{2}\alpha\kappa^2\sigma^2}\frac{_2F_1\left(0,\frac{1}{2};\frac{3}{2};\frac{(\alpha-\frac{\mu^2}{2\alpha})^2}{(\alpha-\frac{\mu^2}{2\alpha})^2+\kappa^2}\right)}{\sqrt{\left(\alpha-\frac{\mu^2}{2\alpha}\right)^2+\kappa^2}} +
     \frac{\mu}{8\sqrt{2}\alpha\kappa^2\sigma^2}.
\end{align}
Eventually, the error probability of the basis scheme is given by
\begin{align}
    &P_{e|\textbf{basis}}= 
    \frac{1}{2}-\frac{\mu}{\sqrt{8}}\times
    \frac{_2F_1\left(0,\frac{1}{2};\frac{3}{2};\frac{\mu^2}{\mu^2+\frac{1}{2\sigma^2}}\right)}{\sqrt{\frac{\mu^2}{2}+\frac{1}{4\sigma^2}}} +\frac{\mu}{8\sqrt{2}\alpha\kappa^2\sigma^2}+\nonumber\\
    &\frac{\mu^3}{16\sqrt{2}\alpha^2\kappa^2\sigma^2}\times
    \frac{_2F_1\left(0,\frac{1}{2};\frac{3}{2};\frac{\mu^4}{\mu^2+4\alpha^2\kappa^2}\right)}{\sqrt{\frac{\mu^4}{4\alpha^2}+\kappa^2}}+\nonumber\\
    & \frac{\mu\left(\alpha-\frac{\mu^2}{2\alpha}\right)}{4\sqrt{2}\alpha\kappa^2\sigma^2}\times \frac{_2F_1\left(0,\frac{1}{2};\frac{3}{2};\frac{(\alpha-\frac{\mu^2}{2\alpha})^2}{(\alpha-\frac{\mu^2}{2\alpha})^2+\kappa^2}\right)}{\sqrt{\left(\alpha-\frac{\mu^2}{2\alpha}\right)^2+\kappa^2}}.
\end{align}